\documentclass[useAMS,usenatbib]{mn2e}

\usepackage[T1]{fontenc}
 \usepackage{pslatex}
\usepackage{amsmath}
\usepackage{amsfonts}
\usepackage{xcolor}
\usepackage{url}
\usepackage{bm}
\usepackage{graphicx}
\usepackage{amssymb}
\usepackage{soul}
\usepackage{booktabs}
\usepackage{array}
\usepackage[utf8]{inputenc}
\inputencoding{latin1}
\inputencoding{utf8}

\def\lsim{ \lower .75ex\hbox{$\sim$} \llap{\raise .27ex \hbox{$<$}} }
\def\gsim{ \lower .75ex \hbox{$\sim$} \llap{\raise .27ex \hbox{$>$}} }

\renewcommand{\vec}[1]{\boldsymbol{#1}}

\def\lsim{ \lower .75ex\hbox{$\sim$} \llap{\raise .27ex \hbox{$<$}} }
\def\gsim{ \lower .75ex \hbox{$\sim$} \llap{\raise .27ex \hbox{$>$}} }
\title[...] 
{Current driven kink instabilities in relativistic jets: \\dissipation properties}

\author[G. Bodo et al.] {G. Bodo$^{1}$\thanks{E-mail:
gianluigi.bodo@inaf.it}, G. Mamatsashvili$^{2,3}$, P. Rossi$^{1}$ and A. Mignone$^{4}$\\
$^{1}$INAF/Osservatorio Astrofisico di Torino, Strada Osservatorio 20, 10025 Pino Torinese, Italy\\
$^{2}$Helmholtz-Zentrum Dresden-Rossendorf, Bautzner Landstraße 400, D-01328 Dresden, Germany\\
$^{3}$Abastumani Astrophysical Observatory, Abastumani 0301, Georgia,\\
$^{4}$Dipartimento di Fisica,  Universit\`a degli Studi di Torino, Via Pietro Giuria 1, 10125 Torino, Italy}

\begin{document}

\maketitle

\begin{abstract} 
We analyze the evolution of current driven kink instabilities of a highly magnetized relativistic plasma column, focusing in particular on its dissipation properties. The instability evolution leads to the formation of thin current sheets where the magnetic energy is dissipated. We find that the total amount of dissipated magnetic energy  is independent of the dissipation properties. Dissipation occurs in two stages: a peak when the instability saturates, which is characterized by the formation of a helicoidal  current sheet at the boundary of the deformed plasma column, followed by a weaker almost flat phase, in which turbulence develops. The detailed properties of these two phases depend on the equilibrium configuration and other parameters, in particular on the steepness of the pitch radial profile, on the presence of an external axial magnetic field and on the amount of magnetization. These results are relevant for high energy astrophysical sources, since current sheets can be the sites of magnetic reconnection where  particles can be accelerated to relativistic energies and give rise to the observed radiation.
\end{abstract}


\section{Introduction}
Relativistic magnetized outflows, mostly in the form of collimated jets, power some of the most energetic,  high energy astrophysical  sources, like radio-loud galaxies, Blazars, gamma-ray bursts (GRB) and microquasars. The jets originate in the vicinity of central compact objects, in most cases a black hole, and carry their energy partly in the form of Poynting flux and partly in the form of kinetic energy flux. Jet energy can then be partly dissipated and channeled to energize a population of non-thermal relativistic particles, that give rise to the observed high energy emission. Understanding the physics and the properties of dissipation regions in relativistic jets is therefore fundamental for interpreting the phenomenology of high energy astrophysical sources.  Two main mechanisms are considered to be at the origin of dissipation, namely collisionless shocks and magnetic reconnection.  However, it has been shown \citep{Sironi15} that, for large magnetization values, such as those expected in jets,  shocks are typically rather inefficient and magnetic reconnection has gained a lot of attention as a viable mechanism for the jet energy dissipation. Reconnection regions can originate through the evolution of magnetohydrodynamic (MHD) instabilities, in particular current driven kink instabilities (but also Kelvin-Helmholtz instabilities, see \citet{Sironi21}). For this reason, the study of this kind of instability has recently gained a lot of attention. Linear studies have been performed both in the non-relativistic \citep{Bodo16} and relativistic \citep{Bodo13, Bodo19, Sobacchi17, Kim17, Kim18} regimes and the nonlinear evolution of the instability has been  studied through numerical simulations \citep{Mizuno09, Bromberg19, Kadowaki21, Medina21}. More recently \citet{Zhang16, Bodo21} have  analyzed the properties, in particular the polarization properties,  of the emitted  radiation, by a simplified model in which emitting particles are assumed to be originated at the current sheets formed during the evolution of the kink instability. Om a smaller scale, the observational signatures of reconnection layers have been studied by means of PIC simulations by \citet{Zhang18, Sironi20}. The same chain of  processes have been studied in a completely different astrophysical context, in a non-relativistic regime,  namely in the context of flares and heating of the solar corona by \citet{Browning03, Browning08, Gordovskyy11, Gordovskyy17}.

The aim of this paper is to analyze in detail the dissipation properties during the nonlinear evolution of the kink instability in jets. Since the main avenue through which dissipation occurs is the formation of thin current sheets, our study will focus on the derivation of their characteristics. In addition, we will characterize the properties of turbulence that develops in the later stages of the instability evolution. \citet{Bromberg19} claim that turbulence is present during the evolution, however they do not give any quantitative measure of its properties, which we instead present in this paper. We also consider the effect of different types of equilibria and of different values of the main parameters, like, for example, the initial magnetization. As in \citet{Bromberg19}, we limit this study to the case of a static, non-rotating plasma column, the effect of a velocity structure will be postponed to a following paper.

In the next Section \ref{sec:simul} we describe our numerical setup, the equilibrium configurations, the parameters that determine these configurations and the simulations that will be analysed in the paper. In Section \ref{sec:res} we present the results starting from a reference case and comparing to it all other cases. Finally, the summary and discussion  are given in Section \ref{sec:disc}.

\section{Numerical setup}
\label{sec:simul}
Our aim is the analysis of the evolution of the current driven kink instability of a highly magnetized, relativistic, plasma column. The relevant equations are those of ideal relativistic MHD:
\begin{equation}
 \partial_t \left( \gamma \rho \right) + \nabla \cdot \left(\gamma \rho \vec{v}\right)= 0,
  \label{eq:drho/dt} \\ 
\end{equation}
\begin{equation}
\partial_t \vec{m} + 
\nabla \cdot 
\left( \gamma^2 w \ \vec {v} \vec {v} - \vec {E} \vec {E} - \vec {B} \vec {B} + (p+u_{\mathrm {em}})\vec{I} \right)= 0,
\label{eq:dm/dt} \\ 
\end{equation}
\begin{equation}
 \partial_t \left( \gamma^2 w -p + u_{\mathrm {em}} \right) + \nabla \cdot \left( \gamma^2 w \vec {v} + \vec {E} \times \vec {B} \right)= 0,
\label{eq:energy}
\end{equation}
\begin{equation} \label{eq:dB/dt}
\partial_t \vec {B} + \nabla \times \vec {E}=0,
\end{equation}
where $\rho$ is the proper density,  $\vec{m} = \gamma^2 w \vec {v} + \vec {E} \times \vec {B}$ is the momentum density, $p$ is the pressure, $w$ is the relativistic enthalpy, $\gamma$ is the Lorentz factor,  $\vec {v}$, $\vec {B}$, $\vec{E}$  are, respectively, the velocity, magnetic field and electric field 3-vectors, $u_{em} = (E^2+B^2)/2$, $\mathbf I$ is the unit $3 \times 3$ tensor and the electric field is provided by the ideal condition  $\vec{E}+\vec{v}\times \vec{B}=0$. Additionally we have to specify an equation of state relating $w$, $\rho$ and $p$.  We will both consider a $\gamma$-law with constant $\Gamma$ and  the Taub-Matthews equation of state \citep{Mignone05}. 
The units are chosen so that the speed of light is $c = 1$, we also remark that a factor of $\sqrt{4 \pi}$ will be reabsorbed in the definitions of $\vec{E}$ and $\vec{B}$.   

\begin{figure*}
\centering%
\includegraphics[width=\textwidth]{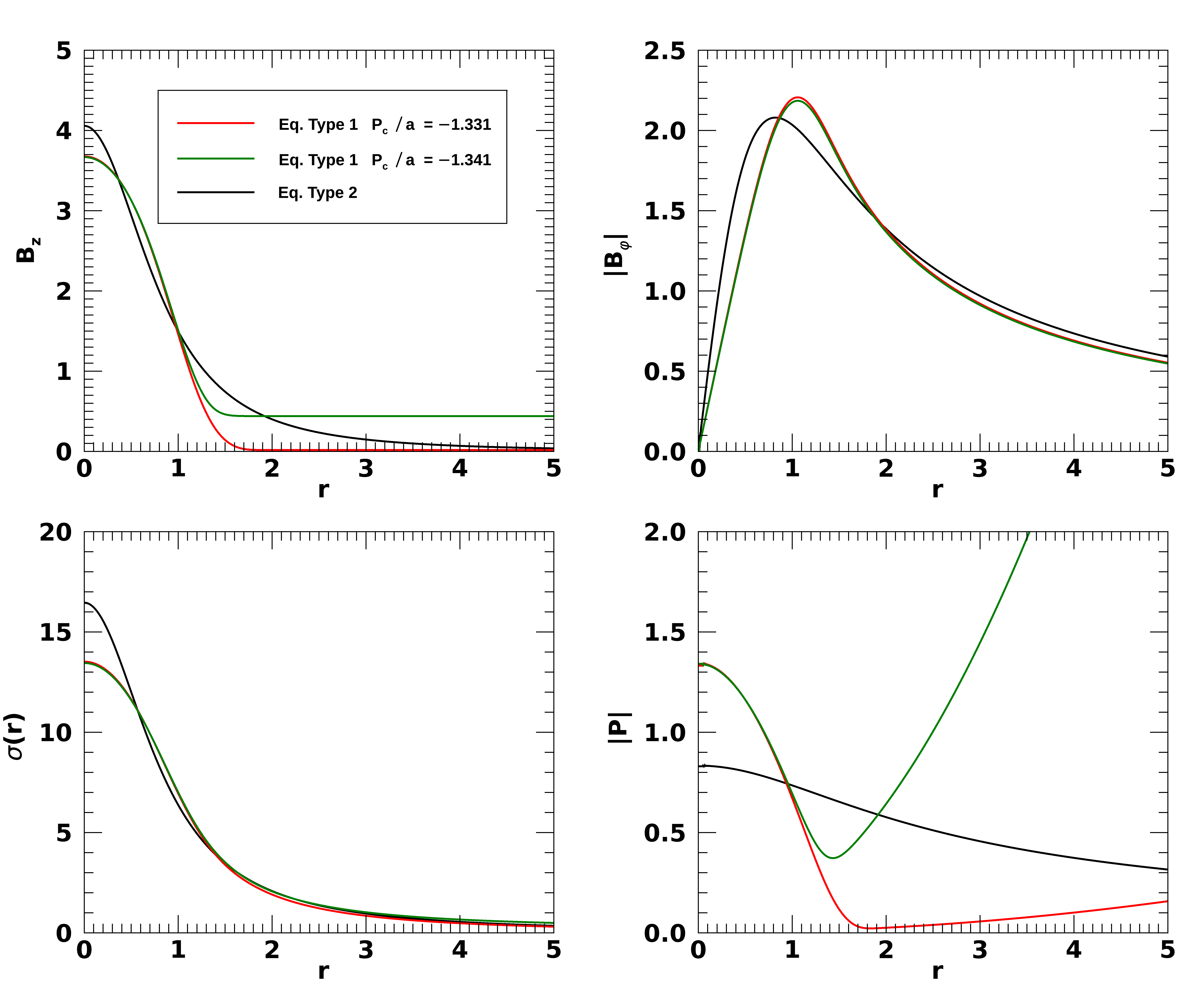}%
\caption{Plots of the equilibrium profiles of $B_z$ (top left panel), $B_\varphi$ (top right panel), local magnetization $<sigma(r)$ (bottom left panel)  and the pitch $|P(r)|$ (bottom right panel) as a function of radius. In each panel, the red and green curves refer to Type I equilibria with two different values of $P_c$, while the black curve refers to Type II equilibrium.}
\label{fig:equil} 
\end{figure*}

At $t=0$, we setup an axisymmetric equilibrium structure with zero velocity, constant density $\rho_0$ and constant pressure, for which, considering cold jets, we take the value  $p_0 = 0.01 \rho_0 c^2$. Since we intend to consider highly magnetized configurations, we will assume a force-free equilibrium: in this case the equilibrium condition reads 
\begin{equation}\label{eq:radial_eq}
  \frac{1}{2r}\frac{d(r^2 B_\varphi^2)}{dr} 
  + \frac{r}{2}\frac{dB_z^2}{dr}  = 0\,.
\end{equation}
Equation (\ref{eq:radial_eq}) leaves the freedom of choosing the radial profile of one of the two components of magnetic field and then solve for the other. In  this work, we mainly consider the equilibrium magnetic field introduced  by \citet{Bodo13} (equilibrium Type I):
\begin{equation} \label{eq:Bphi_prof}
 B_\varphi^2 = \frac{B_0^2}{(r/a)^2}\left[1 - \exp\left(-\frac{r^4}{a^4}\right)\right]\, ,
\end{equation}
\begin{equation}\label{eq:Bz_prof}
  B^2_z = B_0^2 \left[\frac{P_c^2}{a^2} -  \sqrt{\pi}\,{\rm erf} \left(\frac{r^2}{a^2}\right) \right]\, ,
\end{equation}
\begin{equation}\label{eq:Br}
    B_r = 0\, ,
\end{equation}
where erf is the error function, $r$ is the cylindrical radius and $a$ is the magnetization radius, i.e. the radius inside which most of the magnetic energy is concentrated. We take negative $B_{\varphi}<0$ and positive $B_z>0$ for this equilibrium, as used in \citet{Bodo13}. We will also make comparisons with the equilibrium proposed by \citet{Mizuno09} (equilibrium Type II), for which the field configuration is
\begin{equation} \label{eq:Bphi_prof_m}
 B_\varphi = - \frac{a B_z}{r}\sqrt{\frac{\left[1+(r/a)\right]^{2 \alpha}-1 - 2 \alpha (r/a)^2}{2 \alpha - 1}} \,,
\end{equation}
\begin{equation}\label{eq:Bz_prof_m}
  B_z = \frac{B_0}{\left[1+(r/a)\right]^{\alpha}} \,,
\end{equation}
\begin{equation}\label{eq:Br_m}
    B_r = 0\,.
\end{equation}
Equilibrium Type I depends on two parameters, the pitch  on the axis defined as
\begin{equation}
    P_c \equiv \left. \frac{r B_z}{B_\varphi} \right|_{r=0}  < 0
\end{equation}
and  the average magnetization of the jet defined as
\begin{equation}
    \sigma = \frac{\langle B^2 \rangle}{\rho_0  c^2} \,,
\end{equation}
 where $ \langle B^2 \rangle$ is  defined as
\begin{equation}
    \langle B^2 \rangle = \frac{\int_0^a (B^2_z + B_\varphi^2) r dr}{\int_0^a r dr} \,
\end{equation}
and  the value of $\sigma$ determines the parameter $B_0$ appearing in equations (\ref{eq:Bphi_prof}) and (\ref{eq:Bz_prof}). This equilibrium has a minimum allowed value of $|P_c|/a$, namely for $|P_c|/a < \pi^{1/4}$ no equilibrium configuration is possible. In the equilibrium Type II, we have the same definition for the magnetization that allows us to similarly determine $B_0$ in equations (\ref{eq:Bphi_prof_m}) and (\ref{eq:Bz_prof_m}), but, instead of $P_c$, we have the parameter $\alpha$ that determines both the value of the pitch on the axis  and its radial dependence. More precisely, for $\alpha < 1$ the pitch  increases with the radius $r$, for $\alpha = 1$ the pitch is constant and for $\alpha > 1$ the pitch decreases with radius. Recently \citet{Bromberg19} have performed a series of simulations of the evolution of the current driven kink instability in jets, comparing different initial equilibria that possessed different pitch profiles and  pointed out the importance of these profiles in determining the evolution of the instability, in particular the case of pitch decreasing with radius appears to be the most efficient and fast in dissipating the magnetic energy. For this reason, we will concentrate on this kind of profiles and, more precisely for the equilibrium Type I, we show the profiles for two  values of $P_c$, namely  $P_c/a = -1.331$ very close to the minimum allowed value, and $P_c/a = -1.341$, while for equilibrium Type II we choose $\alpha = 1.44$. In Fig. \ref{fig:equil} we compare the profiles of $B_\varphi$, $B_z$ and the pitch for these three equilibria with the same magnetization $\sigma = 10$. We can see that, for the equilibrium Type I,  $B_z$ decreases faster, reaching, for $r/a \gsim{3}$ a constant, very small value.  The behaviour of the toroidal field is instead quite similar for all the  cases. The different behaviour of $B_z$ translates in a different pitch profile, the central pitch is lower (in absolute value) for Type II, but Type I has a much faster decrease reaching at $r/a \sim 1.8$ the minimum value very close to zero. In addition, in the same figure, we also show the profiles for an equilibrium Type I, but with a higher value of $P_c/a = -1.341$ (green curve). In this case the constant value of $B_z$ reached for $r \gsim{3}$ in much larger than in the previous case (red curve). In the bottom left panel we show the radial behavior of the local magnetization, which is higher on the axis for the equilibrium Type I, but shows a faster decrease with radius.

The system of equations (\ref{eq:drho/dt})-(\ref{eq:dB/dt}) is set in dimensionless by using the magnetization radius $a$ as the unit of length, the light crossing time over the magnetization radius as the unit of time and the initial uniform density $\rho_0$ as the unit of density. Together with equations (\ref{eq:drho/dt})-(\ref{eq:dB/dt}),  we also consider the evolution equation for a passive tracer $f$:
\begin{equation}
\frac{\partial}{\partial t} \left( \gamma \rho f\right) +\nabla \cdot \left(  \gamma \rho f \vec{v}\right)= 0 \,.
\end{equation}
The passive tracer $f$ is set to 1 inside the magnetization radius and 0 outside and it will allow to study the mixing of the material inside the magnetized region with the external material.  In fact, during the evolution, $f$ will take values between 0 and 1 as a results of the mixing process.

\begin{table*}\label{tab:casi} 
\centering
\begin{tabular}{lrrrcrcccc}\hline                                        
Case    & $\sigma$ &  $P_c$   & $L_z$    & EOS                        & Equilibrium   &  L  & $t_f$  &  $N_x \times N_y \times N_z$   \\
\hline\hline\noalign{\medskip}
 Ref            &  $10$    &    $-1.331$       & $16.67$     & $\Gamma$-law   & Type I  & $100$  &  $500$   &  $1400\times 1400\times250$~~(LR)  \\
 RefHR      &  $10$      & $-1.331$         & $16.67$     & $\Gamma$-law   & Type I             & $50$  &  $250$   &  $1400\times 1400\times500$~~(HR) \\
 Ref2L       &  $10$      &  $-1.331$         & $33.33$     & $\Gamma$-law   &  Type I             & $50$  &  $250$   &  $700\times700\times500$~~(LR)  \\
  PitchHi            &  $10$    &    $-1.341$       & $16.67$     & $\Gamma$-law   & Type I             & $50$  &  $250$   &  $700\times 700\times250$~~(LR) \\
 Sigma2.5    &  $2.5$     & $-1.331$            & $16.67$     & $\Gamma$-law   & Type I              & $50$  &  $250$   &  $700\times700\times250$~~(LR) \\
 Sigma1    & $1$      & $-1.331$            & $16.67$     & $\Gamma$-law   & Type I              & $50$  &  $250$   &  $700\times700\times250$ ~~(LR) \\ 
 Taub         & $10$   & $-1.331$            & $16.67$     & Taub                       & Type I               & $50$  &  $250$   &  $700\times700\times250$~~(LR) \\ 
 Eq2           & $10$    &  -         & $16.67$     & $\Gamma$-law    & Type II         & $50$  &  $250$   &  $700\times700\times250$~~(LR)\\ 
 \noalign{\medskip}
\hline
\medskip
\end{tabular}
\caption[]{List of the simulations and parameter sets used in each of the simulations. The first column labels the simulations, the second column gives the magnetization $\sigma$, the third column gives the pitch $P_c$, the fourth column gives the length $L_z$ of the computational domain in the $z$ direction, the fifth, sixth and seventh columns indicate, respectively, the type of equation of state, the equilibria and the lateral extension $L$ of the domain, the eighth column gives the final time of the simulation $t_f$ and the last column gives the number of grid points with the indication of high (HR) or low (LR) resolution runs.}
\label{parset}
\end{table*}

At $t=0$ we perturb the system with a small radial velocity of the form
\begin{equation}
v_r = \epsilon\, r\, \exp(-r^4) \sum_{n=1}^{N} \sin \left(  \frac{2 \pi n z}{L_z} + \phi_n\right) \,,
\end{equation}
where $\epsilon$ is the amplitude of the perturbation (we take $\epsilon = 0.01$), $N$ (we use $N=25$)  is the number of modes that have different longitudinal wavenumbers and $\phi_n$ are random phases,

The simulations are performed on  a Cartesian domain with coordinates in the range  $x\in [-L,L]$, $y\in [-L,L]$ and $z\in [0,L_z]$, which is covered by a grid of $N_x \times N_y  \times N_z$  cells, uniform in the central region around the axis of the magnetized column ($-16 <x <16$ and $-16 < y < 16$) and geometrically stretched towards the boundaries in the $x$ and $y$ directions. 

We performed a set of simulations with different jet configurations in order to explore the role played by different resolutions, different domain length, different initial magnetization, different initial equilibrium configurations and different equations of state. In Table I, we list the parameters and configuration details for all the simulations done. The resolution for the LR simulations is 15 points over the magnetization radius and 30 points for the HR simulations.  We also notice that the reference simulation Ref has a grid that extends laterally up to $L=100$ and a final time $t_f=500$ while all the others have $L=50$ and $t_f = 250$.

The simulations were performed with the PLUTO code \citep{PLUTO}, with parabolic reconstruction, HLL Riemann solver and constrained transport method for the control of the $\nabla \cdot \vec{B} = 0$ condition. The boundary conditions are standard outflow in the $x-$ and $y$-directions, as implemented in the code, and periodic in the axial $z$-direction.

\begin{figure*}
\centering%
\hspace{-1cm}
\includegraphics[width=9cm]{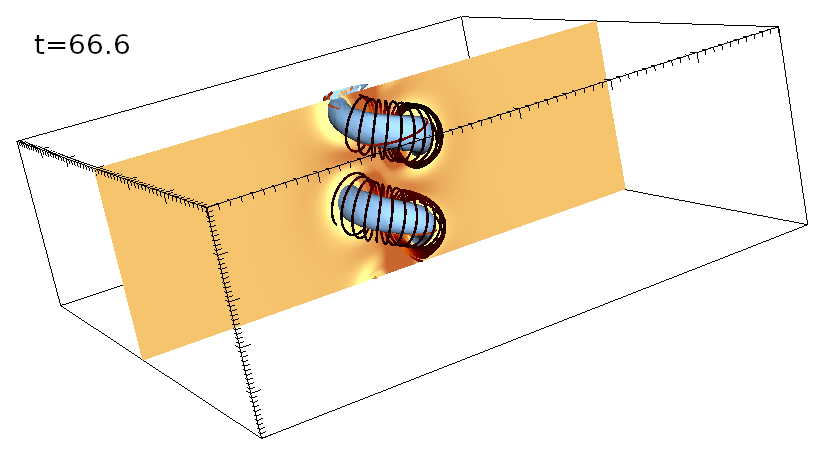} 
\includegraphics[width=9cm]{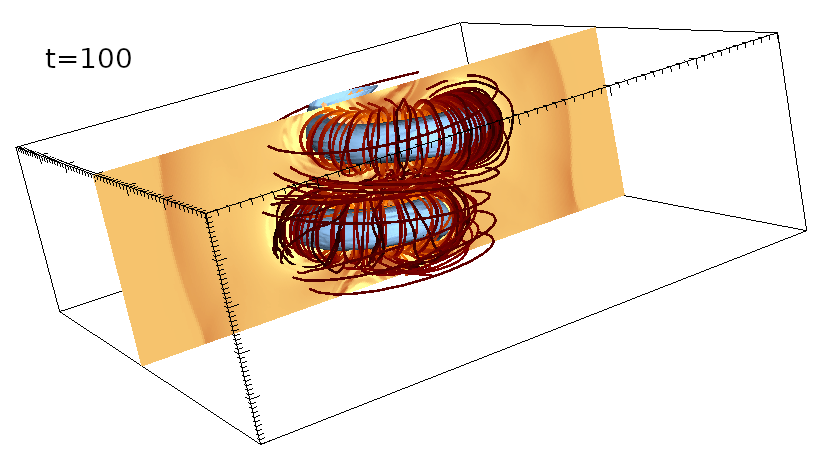} \\
\vspace{0.5cm}
\hspace{-1cm}
\includegraphics[width=9cm]{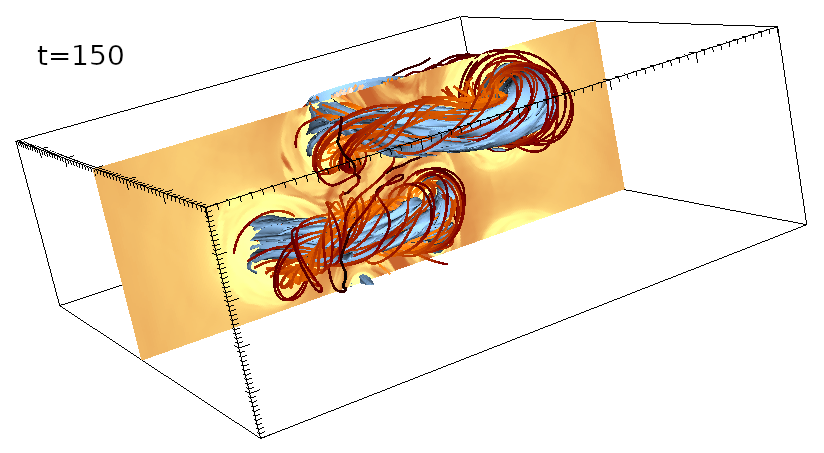}
\includegraphics[width=9cm]{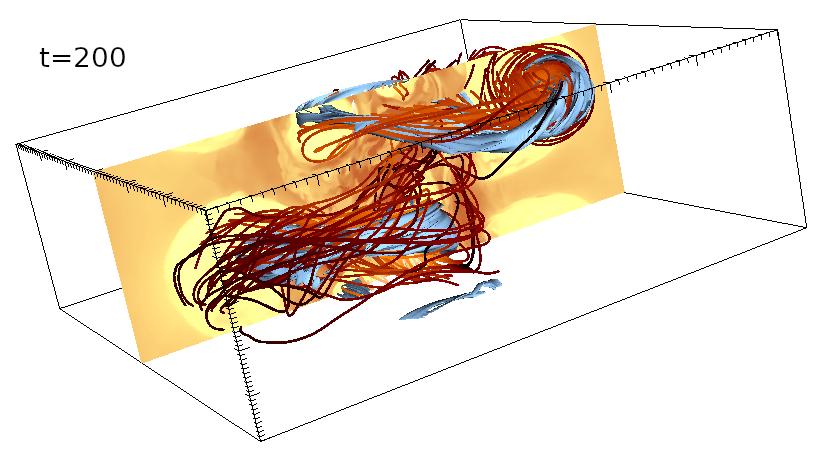}\\
\vspace{0.5cm}
\includegraphics[width=18cm]{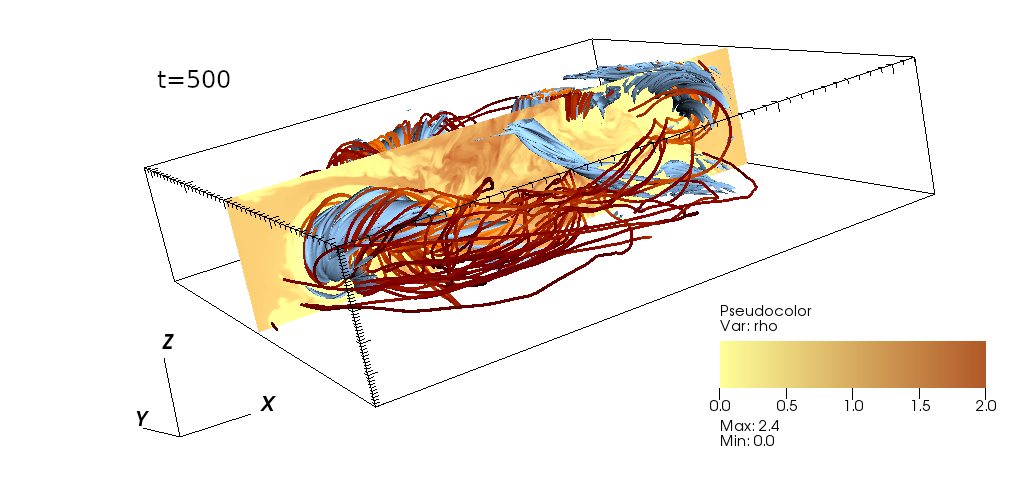}
\caption{Evolution of the kink instability from the linear growth to nonlinear saturation and turbulence. In each panel, we display a three-dimensional isocontour of the tracer distribution (in light blue) with representative magnetic field lines in red and a cut of the density distribution in the $x-z$ plane at $y=0$. The five panels refer to five  different times, respectively $t=66.6$ (top left panel), $t=100$ (top right panel), $t=150$ (left middle panel), $t=200$ (right middle panel and $t=500$ (bottom panel). The boxes in the top four panels extend from $-25$ to $25$ in the $x$ and $y$ directions while the box in the bottom panel extends from $-37$  to $37$. }
\label{fig:comp} 
\end{figure*}

\section{Results}
\label{sec:res}

\begin{figure}
\centering%
\includegraphics[width=8cm]{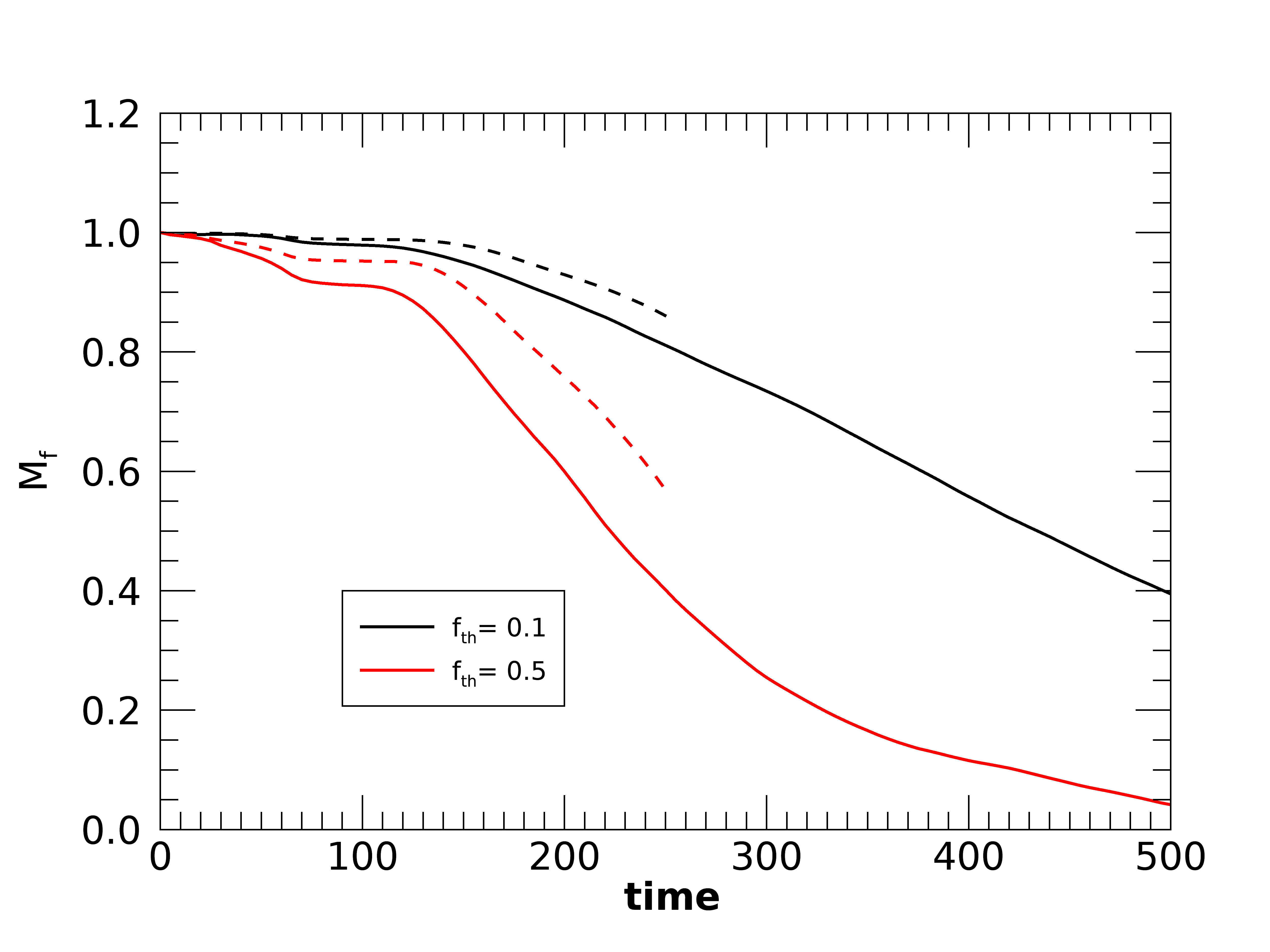}%
\caption{Plot of the tracer mass fraction  of the initial tracer mass, for $f > f_{th}$, as a function of time. We consider two different thresholds, namely $f_{th} = 0.1$ (black curves) and $f_{th} = 0.5$ (red curves). The solid curves refer to the low resolution case (LR), while the dashed curves refer to the high resolution case (HR), for which the final time of the simulation is $t_f = 250$.}
\label{fig:mass} 
\end{figure}

\begin{figure}
\centering%
\includegraphics[width=8cm]{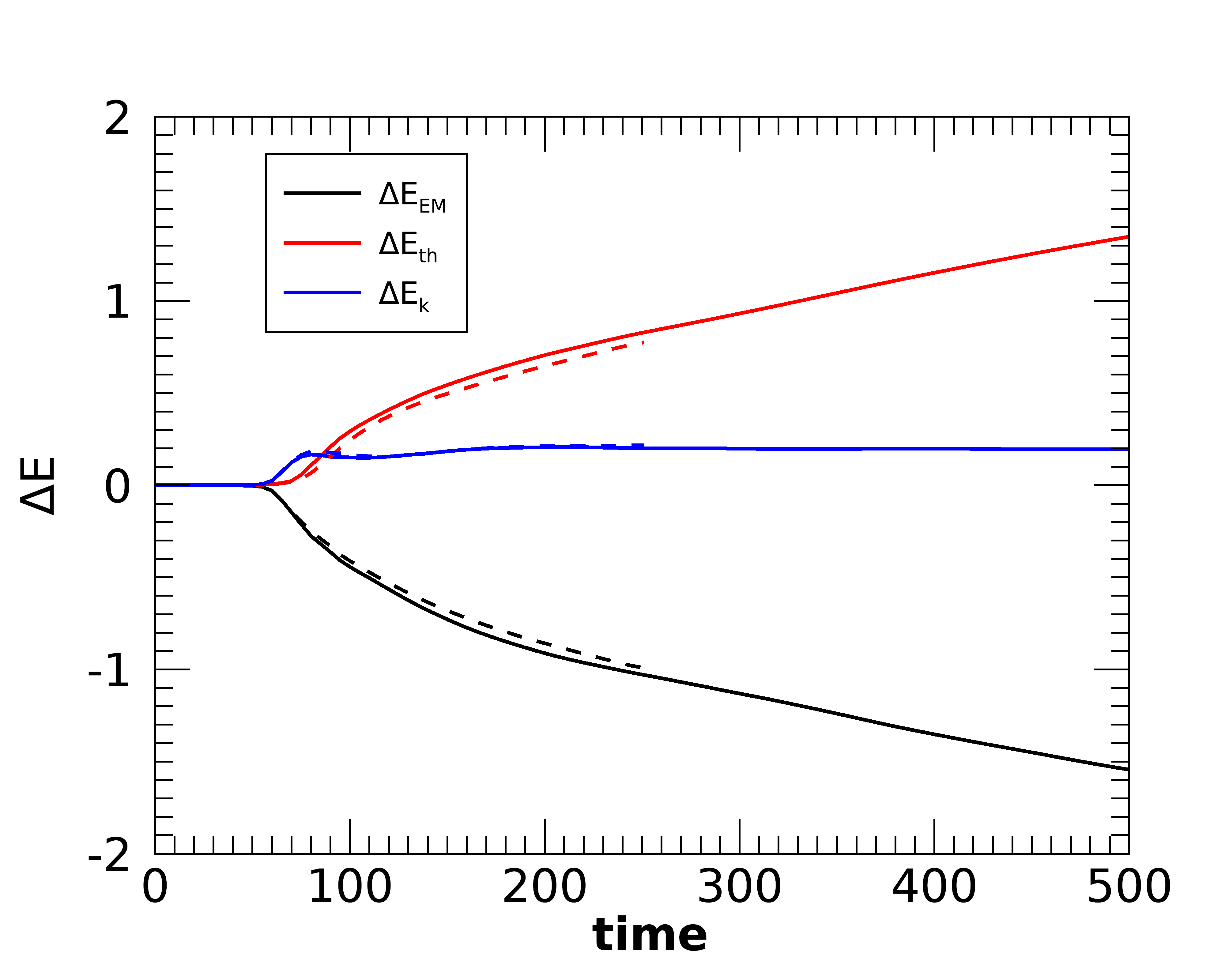}%
\caption{Plot of the variation of the three forms of energy (kinetic energy, thermal energy, electromagnetic energy), as a function of time. The energy variations are shown with respect to the value at $t=0$ and are normalized to the magnetic energy contained inside the magnetization radius $a$ at $t=0$. The solid curves refer to the low resolution case (LR), while the dashed curves refer to the high resolution case (HR), for which the final time of the simulation is $t_f = 250$.}
\label{fig:energies} 
\end{figure}

We will start our discussion with a detailed analysis of the reference cases Ref and RefHR, in these two cases we consider the equilibrium of Type I, a constant $\Gamma=5/3$ equation of state, a magnetization $\sigma = 10$ and the two cases differ for the resolution, the lateral grid extension and the final time of the simulation. We will discuss in detail the instability evolution, the dissipation properties and the development of turbulence, considering the differences caused by the increase in resolution.  We will then make comparisons with all the other cases with different configurations.

\subsection{The reference case}

\subsubsection{Instability evolution}
\citet{Bodo13} have shown that the growth rate ${\rm Im}(\omega)$ of the instability  in the linear regime scales as
\begin{equation}
    {\rm Im}(\omega) \sim \frac{v_A}{P_c} \left( \frac{a}{P_c} \right)^2 f(k_z P_c) \,,
\end{equation}
where $k_z = 2 \pi n/ L_z$, with $n$ integer, is the longitudinal wavenumber of the mode and $v_A$ is the Alfv\'en velocity which, in the relativistic case is defined as function of $\sigma$ as
\begin{equation}
    v_A^2 = \frac{\sigma}{1 + \sigma} \,.
\end{equation}
The function $f(k P_c)$ is independent from $P_c/a$ and is a growing function of $k P_c$ up to $k P_c \sim 1$ where the growth rate drops to zero, the modes with a larger $k$ are then stable. The linear growth rate of the instability therefore increases as we increase $\sigma$ until it saturates for large values of $\sigma$ and also increases as we decrease $|P_c|$. In the present case,  $P_c$ is very close to the minimum allowed value and the maximum growth rate is obtaned for $n = 2$.

\begin{figure}
\centering%
\includegraphics[width=8cm]{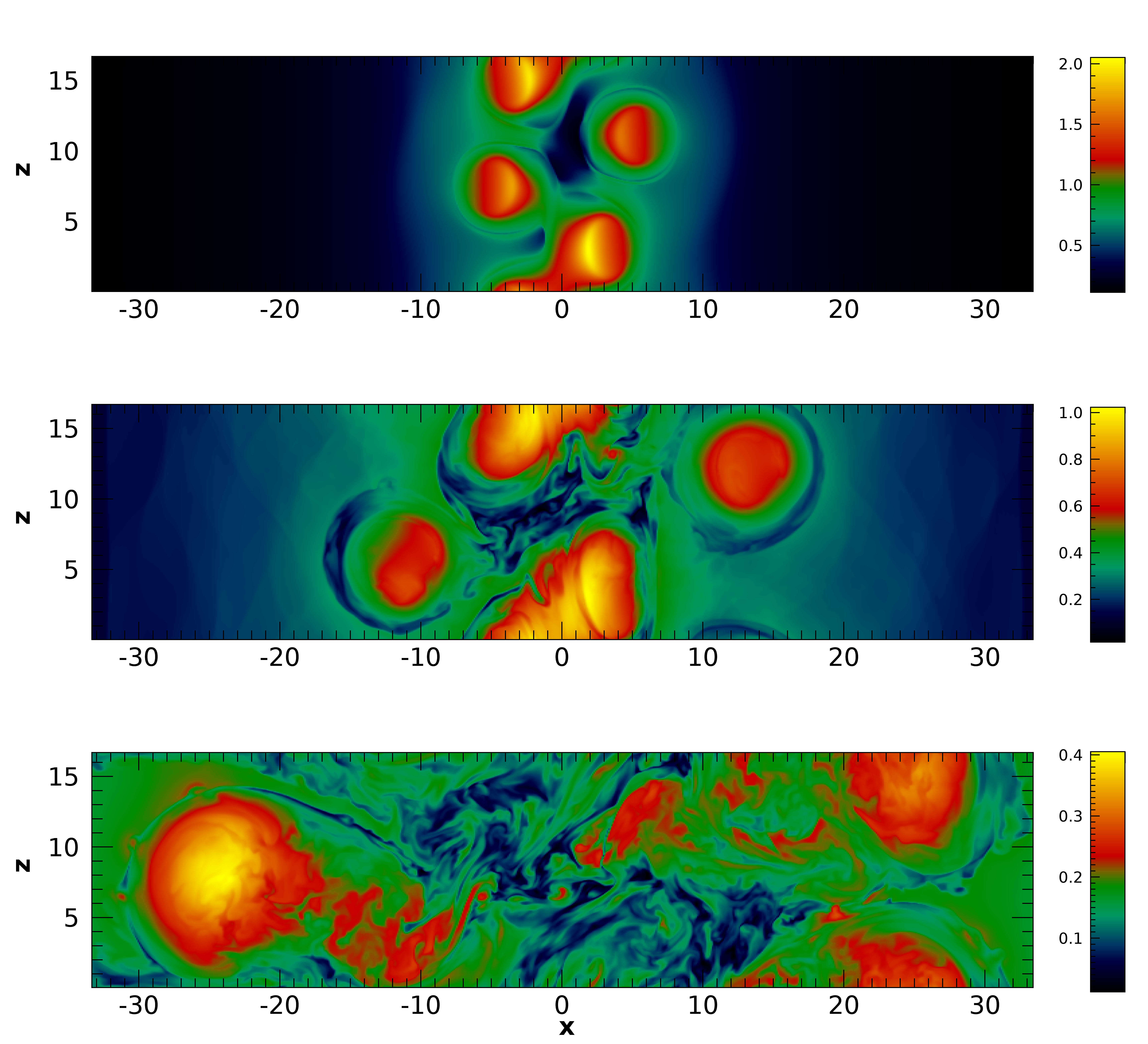}%
\caption{Transverse cuts, in the $x-z$ plane, of the magnetic field intensity at three different times, $t=75$ (top panel), $t=150$ (middle panel) and $t=500$ (bottom panel).}
\label{fig:bfield} 
\end{figure}

In Fig. \ref{fig:comp} we illustrate the instability evolution by showing  three-dimensional composite views of the jet at four different times, $t=66.6$ (top left panel), $t=100$ (top right panel), $t=150$ (middle left panel), $t=200$ (middle right panel) and $t=500$ (bottom panel). Each panel shows in light blue an isosurface of the tracer distribution, a two-dimensional section of the density distribution in the $x-z$ plane and a set of representative magnetic field lines. At $t=66.6$ we observe the formation of the helical deformation of the magnetized column, the field lines appear to wrap around the deformed column.  As predicted by the linear analysis, we observe that the increasing perturbation has a wavenumber $k_z$ corresponding to $n=2$. The density decreases in the region around the deformed column, while it shows an increase in the central part of the domain. At  later times, the deformation increases, the tracer contours become more corrugated, while the magnetic field lines tend to become more aligned  to the helicoidal shape of the tracer contours. We point out that we decreased the value of the tracer isocontour, with increasing time (more precisely it is $0.8$ at $t=66.6, 100$, $0.7$ at $t=150$, $0.5$ at $t=200$ and $0.05$ at $t=500$). The more corrugated shapes of the isocontours and their decreasing values indicate the occurrence of mixing.  A more quantitative measure of the mixing is presented in Fig. \ref{fig:mass}, where we show the fraction $M_f(t)/M_f(0)$, where $M_f$ is the total  mass for the tracer  larger than a given threshold, as a function of time, for simulations Ref (solid curves) and RefHR (dashed curves). More precisely, 
\begin{equation}
M_f (t, f>f_{th}) = \int_{f > f_{th}}   \gamma \rho f  dV \,,
\end{equation}
where   the contributions to the integral come only from  the cells that have a value of the tracer $f$ larger than the threshold $f_{th}$.
In the figure we show two curves for two different values of the threshold, i.e. $f_{th} = 0.1$ and $f_{th} = 0.5$. For interpreting this figure we remind that, at $t=0$, $f$ is set to 1 for  $r<a$ and to 0 outside. As the evolution proceed, as a consequence of mixing, we will observe the appearance of tracer values between 0 and 1. The value of the tracer in a given cell indicates the percentage of jet material that is present in that cell. At the beginning of the evolution, the sharp transition at the jet interface is smoothed out by numerical diffusion and therefore there will be a decrease of $M_f$ more pronounced for  the higher value of the threshold ($f_{th} = 0.5$). The dashed curves refer to a simulation with double resolution, that has lower numerical dissipation, for which the decrease of $M_f$ is less pronounced, as expected.  At later times, instead, the slopes of the solid and dashed curves appear to be quite similar, indicating that mixing occurs in a similar way in both simulations, resulting therefore independent of the (numerical) diffusivity. We can attribute this behavior to the development of  turbulent mixing, where the mixing is due to small scale disordered motions. We will return later to a more detailed characterization of the turbulence that develops during the kink instability evolution.  

\begin{figure}
\centering%
\includegraphics[width=8cm]{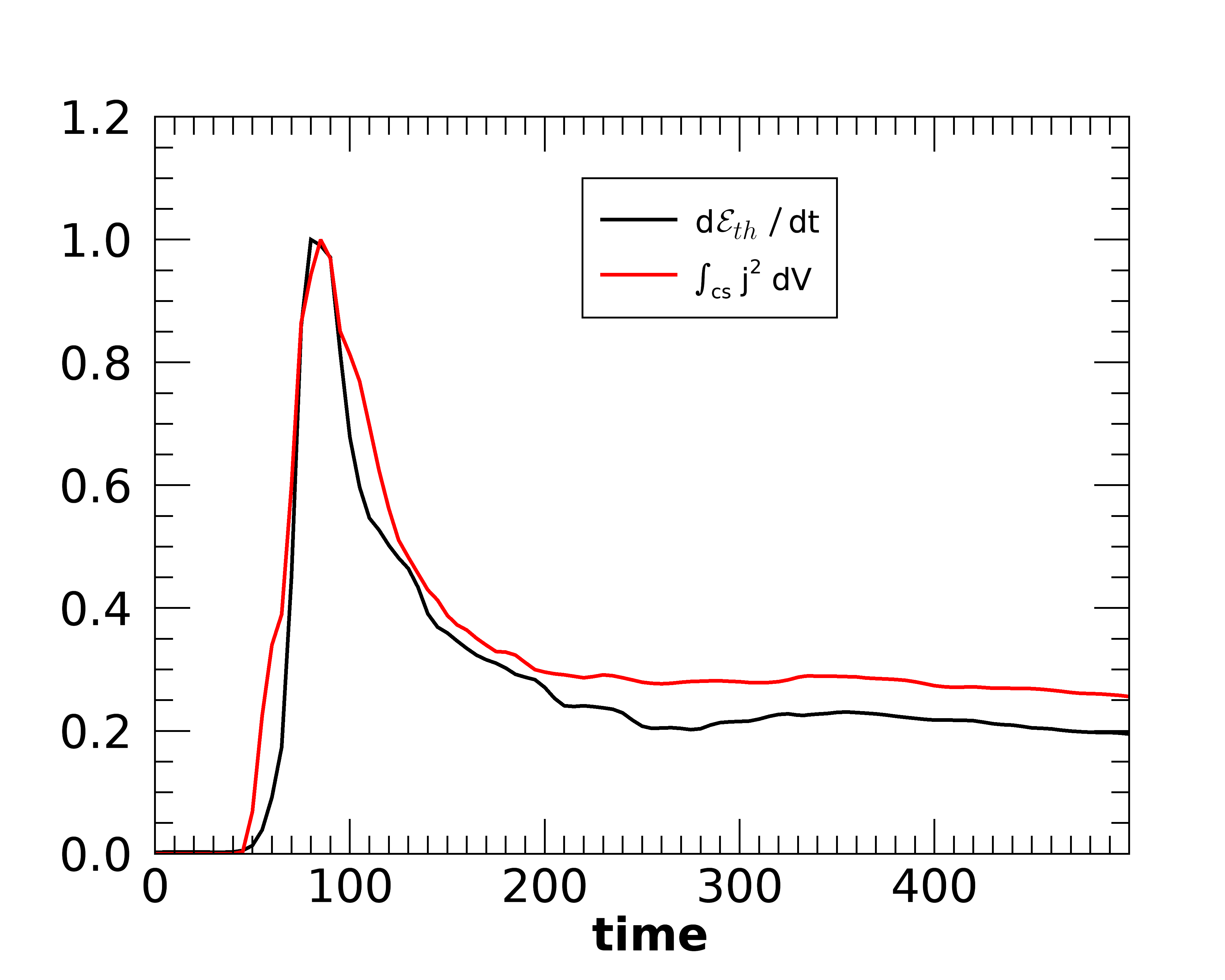}%
\caption{Plot of the energy dissipation rate (black curve), defined as the time derivative of the integral of the thermal energy over the computational box, ${\cal E}_{th} = \int E_{th} dV$, as a function of time. The red curve represents the integral of $j^2$ in current sheets as a function of time.}
\label{fig:deth_j2} 
\end{figure}

The result of the instability evolution is the formation of dissipative structures that will be analysed in more detail in the following sections. Here we will analyze the energy conversion processes from a global point of view, for this we split the  total energy in three parts,  the electromagnetic energy
\begin{equation}
E_{EM} = \frac{1}{2} (B^2 + E^2),
\end{equation}
a second part, 
\begin{equation}
E_k = \rho \gamma (\gamma-1),
\end{equation}
which in the non-relativistic limit reduces to the classical kinetic energy and a third part, 
\begin{equation}
E_{th} = \rho \gamma^2 (h-1) - p,
\end{equation}
where $h$ is the relativistic specific enthalpy, which in the non-relativistic limit reduces to the classical thermal energy.
In Fig. \ref{fig:energies} we plot the variation,  with respect to their initial values,  of  the integrals  over the computational domain of $E_{EM}$, $E_k$ and $E_{th}$, as functions of time, for the two simulations Ref (solid curves) and RefHR (dashed curves). These variations are normalized to the initial magnetic energy inside the magnetization radius ($E_{Mj}$). Notice that, since the jet is mildly relativistic, $E_{EM}$ is dominated by the magnetic energy. We see that the electromagnetic energy decreases with time  and it is converted partly into thermal energy and partly into kinetic energy.  The kinetic energy after an initial increase stays constant at a level of about 20\%  of $E_{Mj}$ while the thermal energy keeps increasing with time. The rate of dissipation of electromagnetic energy  into thermal energy peaks between $t=65$ and $t=140$ (see also Fig. \ref{fig:deth_j2}), but the process continues at a lower rate also at later times. At $t=500$, the final time of our simulation we observe the conversion of about $1.5 E_{Mj}$ of electromagnetic energy into about $1.3 E_{Mj}$ of thermal energy and $0.2 E_{Mj}$ of kinetic energy. The dashed curves represent the results for the high resolution case, with lower numerical dissipation. We can see that the difference between  the two curves is very small, leading to the conclusion that the dissipation rate is determined mainly by the large scale structure of the flow. 

Fig. \ref{fig:bfield} illustrates the evolution of the magnetic field structure by showing transverse cuts in the $x-z$ plane (at $y=0$) of the distribution of the magnetic field intensity at three different times: during the peak of dissipation at $t=75$ (top panel), at the beginning of the slower dissipation phase at $t=150$ (middle panel) and in the turbulence phase near the end of the simulation at $t=500$ (bottom panel). This figure clearly shows the presence of a helical structure, of which we see the intersections of with the cutting plane at $y=0$, is seen to be present at all time, with a growing radius, that reaches a value of about 25 at the final time. In addition we can see that while at the beginning the helix shows two windings, at the end there is only one winding. Inside the helix we can observe the formation of a more disordered or turbulent structure of the magnetic field.

An interesting question is what is the long term relaxed state reached by the system. A useful approach is provided by Taylor's hypothesis \citep{Taylor74} which postulates that, in a highly conducting plasma, magnetic helicity dissipation is much weaker than energy dissipation \citep{Berger84, Taylor86} and therefore the relaxed state will be that of minimal energy under the constraint of helicity conservation. This state is a linear force-free state, i.e. a state with $\vec{J} = \alpha \vec{B}$, with $\alpha$ constant. The theory has been developed for non-relativistic plasmas, while in this case we are in a relativistic regime at high magnetization. However \citet{Bromberg19}  show that at late times the state reached by the system is anyway close to a Taylor's state. In our case the simulation reaches a final time $t_f = 500$, while in  \citet{Bromberg19} it extends longer up to $t_f = 1760$. So, our simulation has not yet reached relaxation: the equilibrium is not force-free: in fact thermal pressure gradients are still present and the plasma $\beta$ is of the order of unity. We notice also that in this state the magnetic field still maintains a helicoidal configuration, as evident in Fig. \ref{fig:comp}. As it will be discussed below, in this case dissipation is predominantly due to turbulence and there is not yet any sign of turbulence decay. This reference simulation is in contrast to our other simulations at different  field configurations presented below. In those runs, even at comparable times, the structure tends to  approach more a force-free state in the sense that the plasma $\beta$ becomes $\ll 1$ as well as the magnetic field configuration appears to be, on average,  axisymmetric and closer to the final state described by Bromberg et al. (2019).

\begin{figure*}
\centering%
\includegraphics[width=\textwidth]{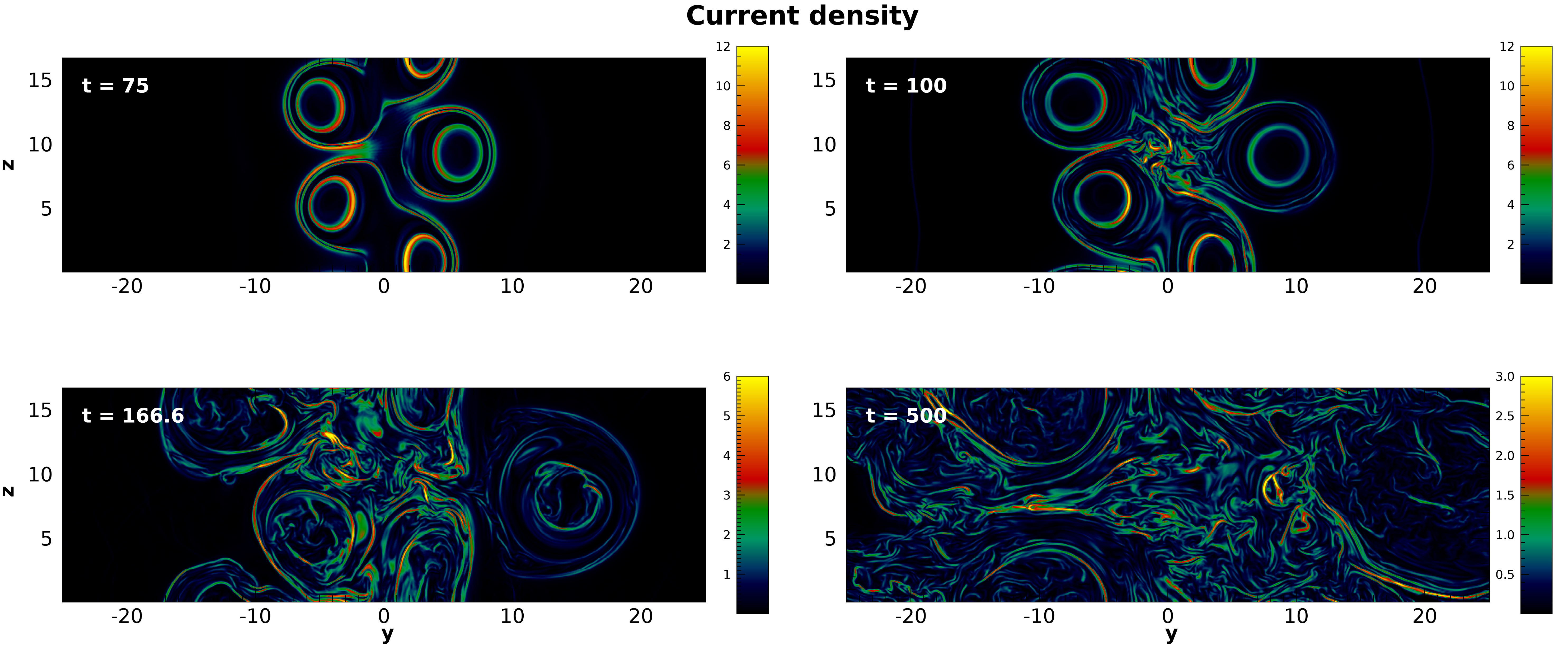} 
\caption{Two-dimensional cuts of the current distribution in the $y-z$ plane at $x=0$. The four panels refer to four different times: $t=75$, top left panel, $t=100$, top right panel, $t=166.6$, bottom left panel and $t=500$, bottom right panel. Notice that the maximum value of the current is different in the different panels.}
\label{fig:current} 
\end{figure*}

\subsubsection{Detection of current sheets}
As we will see in the next subsection, current sheets play a fundamental role in the dissipation process, therefore we have to find a way to identify them. Several different algorithms For their identification and characterization have been proposed,  in particular  \citet{Zhdankin13} proposed a detection algorithm  that first selects points of maximum of the current above some predefined threshold, then builds clusters of points around these maxima in which the current is above half of the maximum.  The threshold is defined as a multiple of the r.m.s. value of the current.  In the present  case the definition of the r.m.s. value is more problematic than in the turbulence case considered by \citet{Zhdankin13} since here the volume of the region in which current sheets are present varies with time, as demonstrated by Fig. \ref{fig:current}. Here  we choose to follow a different approach, which has been already presented in \citet{Bodo21}. Taking into account that we rely on numerical dissipation, we identify as dissipative structures those defined on a small number of grid points, for which numerical dissipation becomes very effective. We then define a local steepness parameter as 
\begin{equation}\label{eq:scale}
    s = \frac{j \delta}{B} \,,
\end{equation}
where $\delta$ is the cell size. $s$ represents a measure of the steepness of magnetic field gradients, the larger is $s$ the smaller is the number of grid points that locally resolve the magnetic field gradient.  We then choose to identify cells belonging to a current sheet as those that have $s$ larger than a certain threshold value $s_{th}$. We have compared our criterion with Zhdankin's algorithm \citep[see][]{Bodo21}  and we find good agreement.

\subsubsection{Dissipation and current sheets}

In resistive simulations,  the energy density dissipation rate is given by $\eta j^2$, where $j$ is the conductive current density \citep[see e.g.][]{Mignone19}. Although our simulations are ideal, we can take $j^2$ as a proxy for the dissipation  rate,  we compute $j$ as $\nabla \times \vec{B}$ neglecting all relativistic corrections, being the velocities at most mildly relativistic.  Dissipation will be actually  concentrated in thin current sheets that naturally form during the evolution of the system. In Fig. \ref{fig:deth_j2} the red curve shows the temporal behavior of the integral of $j^2$ over all current sheets in the computational domain (the definition and detection criterion of current sheets will be discussed below),  the black curve instead shows the behavior of the time derivative of ${\cal E}_{th}$, which is the integral of the thermal energy density $E_{th}$ over the simulation box, which is related to the dissipation rate. Both curves are normalized to their maximum value. Note a very good correlation between these two curves, implying that $j^2$ can be a good measure of the dissipation rate. Moreover we can distinguish, at earlier times,  a peak in the dissipation rate followed, at later times by a phase in which the rate is almost constant. This peak can be related to the formation of strong current sheets which are typically located at the boundaries of the helical deformation, while the second phase can be related to the progressive development of turbulence that will be discussed in more detail in the next subsection.

The formation of thin current sheets during the evolution of the system is demonstrated by Fig.  \ref{fig:current} in which we show a section of  the distribution of the current density in the $y-z$ plane, at four different times.  In the four panel we can recognize the different phases of the kink instability evolution, that we already showed in Fig. \ref{fig:comp} and we clearly see that the current density becomes concentrated in very thin structures. The current sheets in the two top panels delineate the helical jet deformation, they are strong and give rise to the peak in the energy dissipation rate, discussed above.  More disordered turbulent structures start to form in the central region of the top right panel and dominate the distribution in the two bottom panels. 

\begin{figure}
\centering%
\includegraphics[width=\columnwidth]{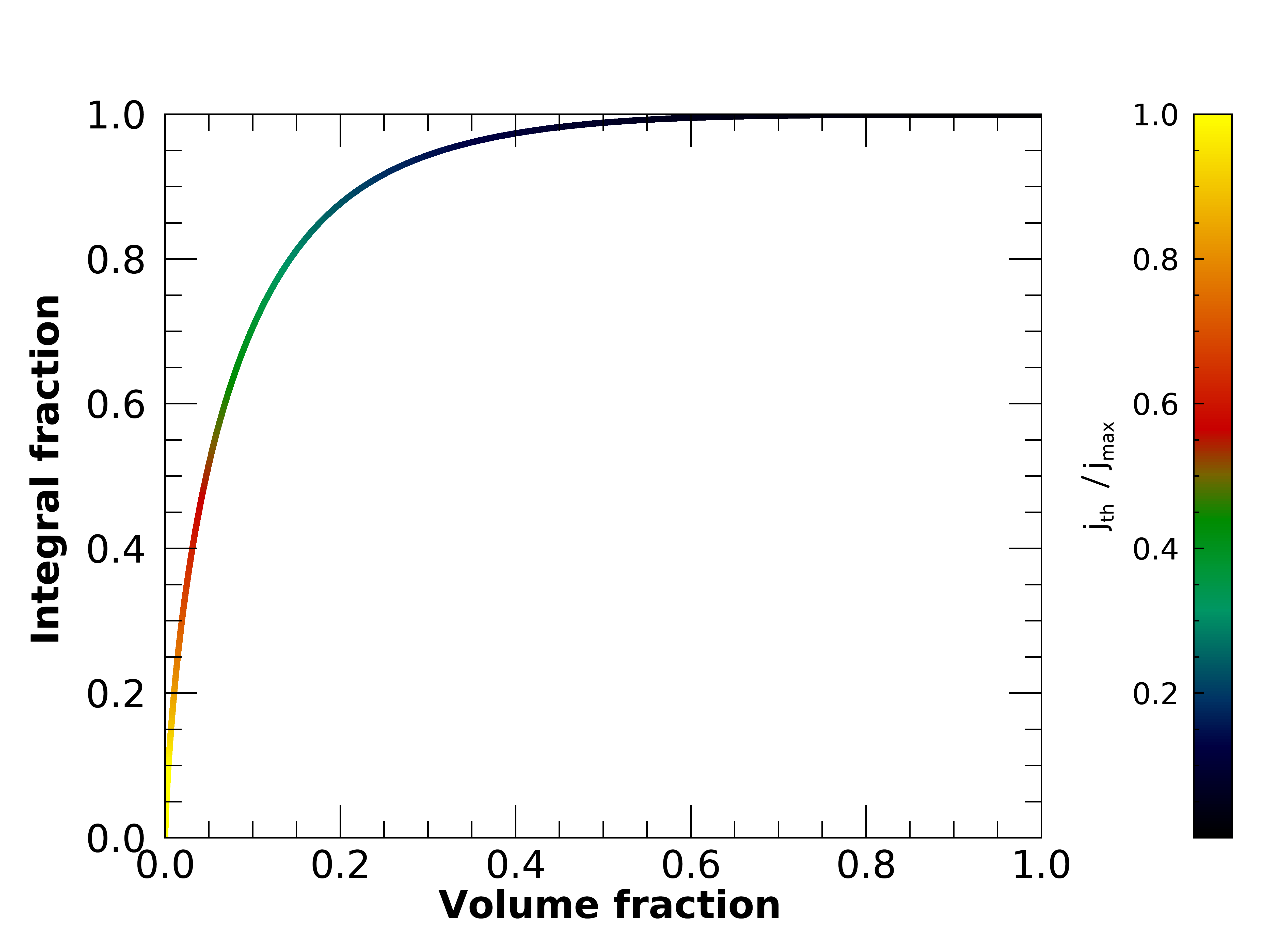}%
\caption{Plot of the fractional contribution to the integral of $j^2$ of regions with $j > j_{th}$ as a function of the volume fraction occupied by the same regions. The curve is coloured with a scale corresponding to the value of $j_{th}/j_{max}$, as indicated by the colorbar. $j_{max}$ is the maximum value of the current in the domain. The plot is for $t=100$. }
\label{fig:cs} 
\end{figure}

\begin{figure*}
\centering%
\includegraphics[width=\textwidth]{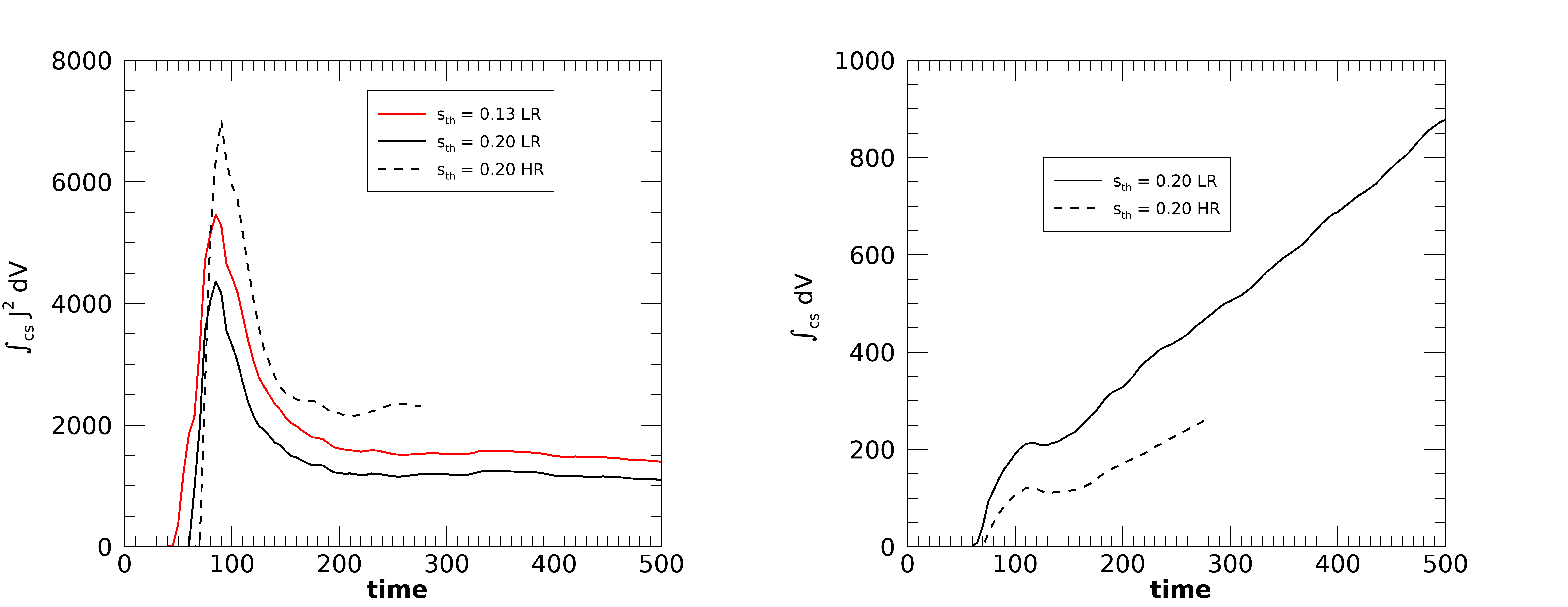}%
\caption{Left panel: Plot of the integral of $j^2$ over the current sheets as a function of time, the different curves refer to different choices of the threshold $s_{th}$ and different resolutions, as specified in the legend. Right panel: Plot of the volume of current sheets as a function of time, for the low (solid) and the high (dashed) resolution runs. Notice that the final time of the HR simulation is $t_f = 250$.}
\label{fig:j2tvol} 
\end{figure*}

\begin{figure}
\centering%
\includegraphics[width=\columnwidth]{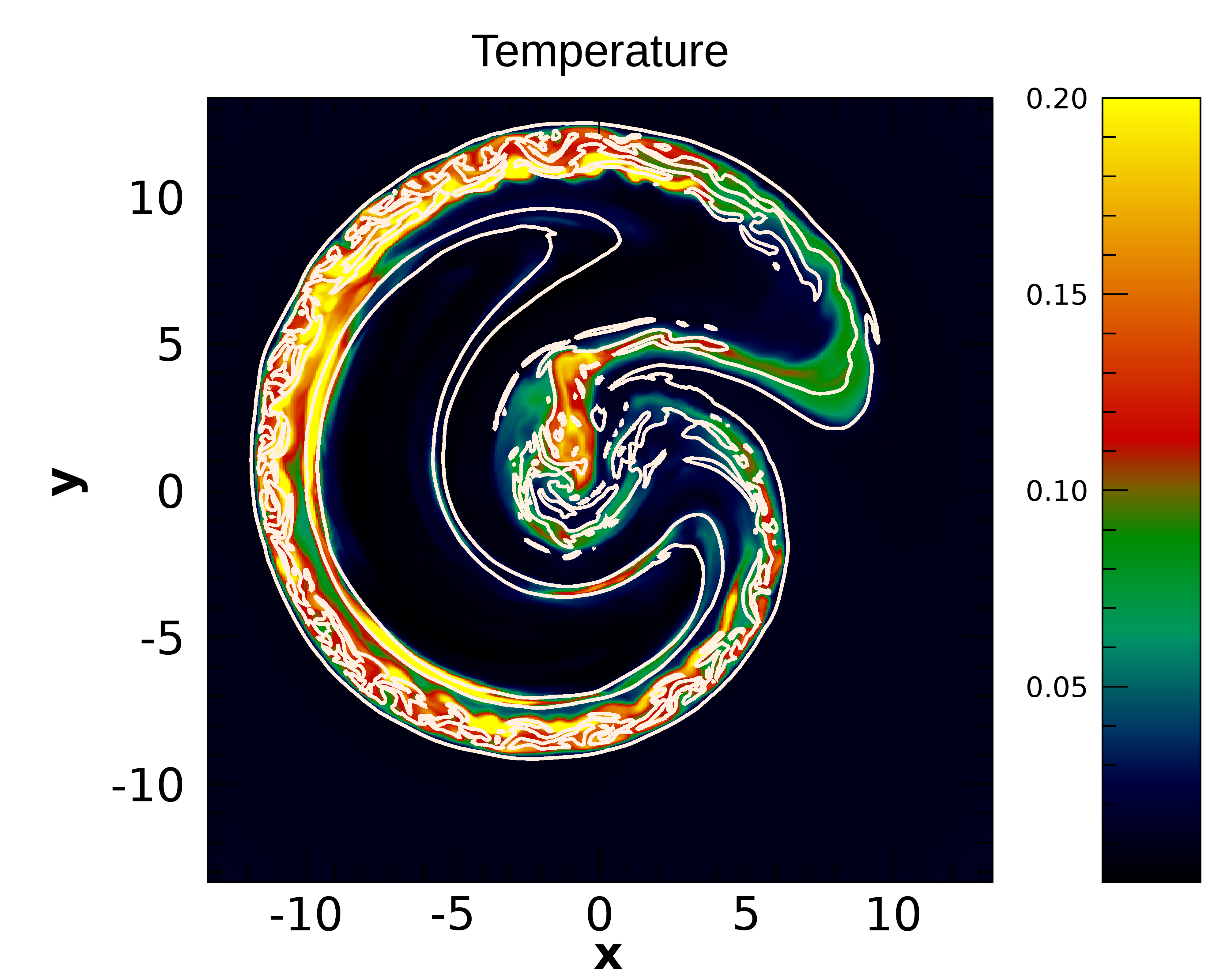}%
\caption{Two-dimensional cut, in the $x-y$ plane, at $z=5$ of the temperature distribution with superimposed a contour plot of $s$ for $s=0.2$. The figure refers to $t=100$. }
\label{fig:temp} 
\end{figure}

The process of concentration of  the current in thin current sheets is more quantitatively illustrated by Fig. \ref{fig:cs} in which we plot the fractional contribution to the integral of $j^2$ from those regions that have a value of $j$ larger than a given threshold current $j_{th}$, as a function of the corresponding  $V(j>j_{th})/V_{ref}$,  i.e., the volume fraction occupied by regions with $j > j_{th}$. The curve in the plot is coloured with a scale that represents the value of $j_{th}$, as indicated in the colorbar. The figure refers to $t=100$, close to the maximum of the dissipation rate (see Fig. \ref{fig:deth_j2}), from Fig. \ref{fig:current} it is seen that at this time the current distribution is concentrated inside a cylinder of radius $\sim 11$, we then take the volume of this cylinder as the reference volume $V_{ref}$. Fig. \ref{fig:cs} shows that if we take a value of $j_{th}$ equal to $0.5 j_{max}$, where $j_{max}$ is the maximum of the current in the domain, the the integral and  volume fractions are, respectively, $\sim 55\%$ and $\sim 7\%$, if we decrease $j_{th}$ to $\sim 0.35 j_{max}$  the corresponding fractions become $\sim 70\%$ for the integral and $\sim 10\%$ for the volume, while for $j_{th} \sim 0.2 j_{max}$, we reach $\sim 90\%$ of the integral in only $\sim 20 \%$ of the volume.

In the left panel of Fig. \ref{fig:j2tvol}, we present the temporal behaviour of the integral of $j^2$ over the current sheets, similar to that shown in Fig. \ref{fig:deth_j2}, except that here we compare different choices of the threshold $s_{th}$ and different resolutions. The black curves refer to the Ref simulation, the dashed curve is for a threshold $s_{th} = 0.13$, corresponding to structures resolved with less than about 8 grid points, while the solid curve is for a threshold $s_{th} = 0.2$, corresponding to a resolution of less than  5 grid points. The behaviour of the two curves is essentially the same and,  as already discussed above, we have an initial peak followed by a flat region for $t>150$, of course with the lower value of the threshold we select more points and the value of the integral is consequently larger, however the relative difference never exceeds $20\%$. The red curve corresponds to the high resolution case RefHR (for a threshold $s_{th} = 0.2$) and a comparison with the corresponding black solid curve shows that again the behaviour is the same, but the values are somewhat less than double. Increasing the resolution,  the transverse  extensions of the current sheets remains the same when measured in grid points, but decreases in size and, correspondingly, the gradients become steeper and the current increases. A more quantitative view of the size variation of the current sheets is provided by the right panel of Fig. \ref{fig:j2tvol}, where we plot the volume of the current sheets as a function of time, the two curves are for the two resolutions: the black curve  is for the lower resolution simulation Ref, while the red curve is for the higher resolution RefHR. In this plot we can see that the volume in the higher resolution case is essentially half of the lower resolution case, while the change in resolution is by a factor of 2. The shape of the current sheets is defined by three dimensions, their length, their width and their thickness. The variation of the volume by only a factor of about 2 seems to imply that only the thickness varies with resolution. At the current peak ($t \sim 80$) we have essentially a single  large scale helicoidal current sheet, for which only its thickness  can vary. More complex is the situation at later times, when turbulence develops, in which, in principle, we could have a variation of all the dimensions, our results, during these later phases, are in agreement with \citet{Makwana15}, who, in a turbulent simulation,  finds a variation of only the thickness of current sheets with resolution.

A further illustration of the relation between current sheets and dissipation is provided by Fig. \ref{fig:temp}, in which we show a two-dimensional cut, in the $x-y$ plane at $z=8.3$, of the temperature distribution at $t=100$, near the maximum of the initial current density peak. Superimposed on the image of the temperature distribution we have plotted a contour of the parameter $s$ for a fixed value of 0.2. These contours, as discussed above, individuate the position of the current sheets. From the figure we notice a strong correlation between the high temperature regions and the position of the current sheets. However, this strong correlation can be observed only at the beginning of the dissipation phase, because the thermal energy does not give an instantaneous image of the dissipation rate, but integrates over time the effect of dissipation, therefore at later times the correlation tends to disappear. Other effects, such as diffusion and advection by the velocity field, tend to destroy the correlation.

\begin{figure}
\centering%
\includegraphics[width=\columnwidth]{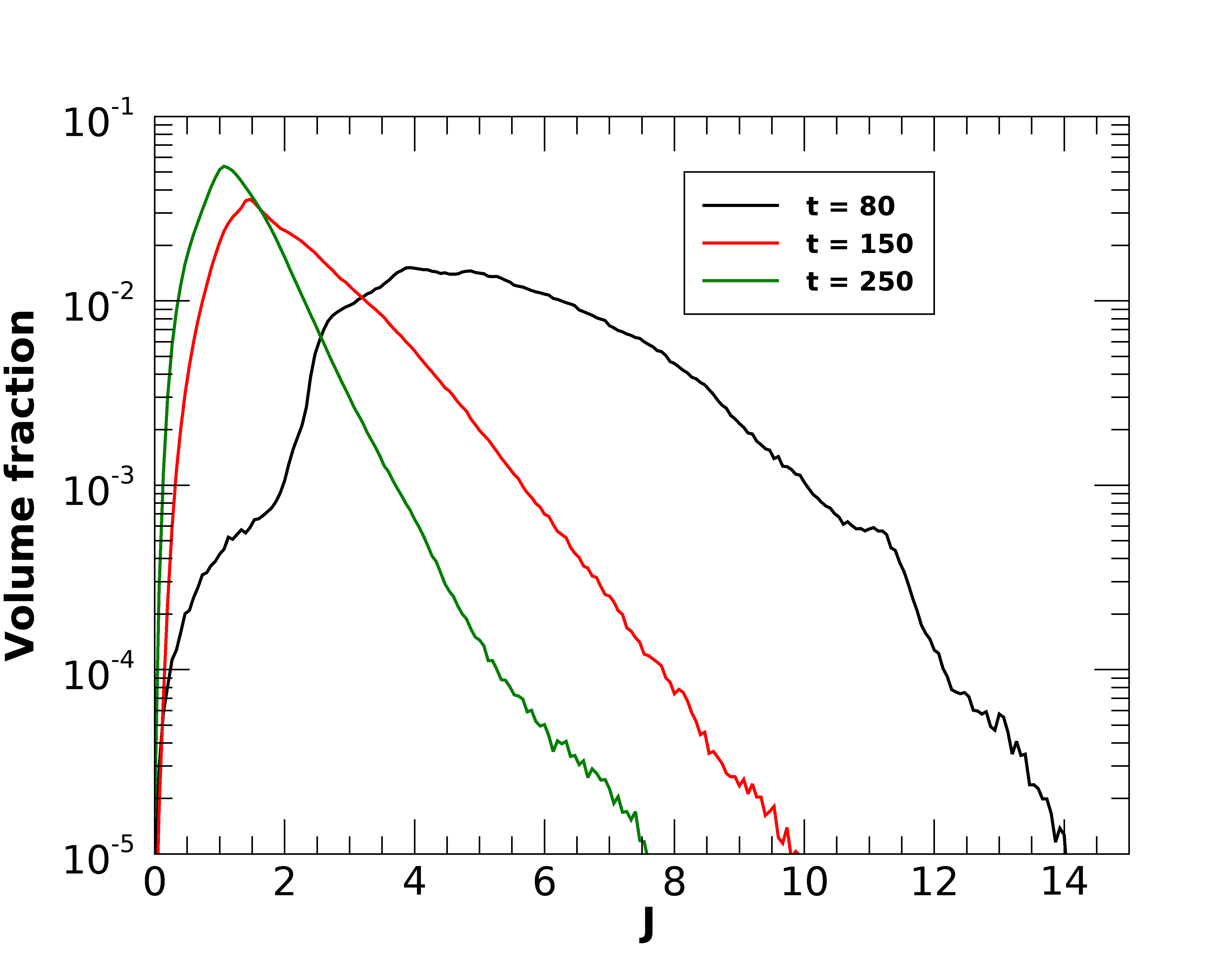}%
\caption{Plot of the volume fraction occupied by current with current density $j$ as a function of $j$ at three different times: $t=80$ (black curve), $t=150$ (red curve) and $t=250$ (green curve).}
\label{fig:jhisto} 
\end{figure}

Fig. \ref{fig:j2tvol} shows that, for $t > 150$, the total current remains constant while the volume of current sheets keeps increasing, this means that the average value of the current has  to decrease with time.  This is confirmed by Fig. \ref{fig:jhisto} where we show the current sheets volume fraction  as a function of current density $j$. In the plot there are three different curves which refer to three different times, $t=80$ (black curve) around the maximum of the total current (see Fig. \ref{fig:j2tvol}), $t=150$  (red curve) at the end of the peak and $t=250$ in the flat phase. We can notice that, not only the peak of the current distribution shifts, with increasing time, towards lower values, but also the shape of the distribution changes with time, showing a fat tail at the first time and progressively decreasing its extension and acquiring an exponential shape.

\begin{figure}
\centering%
\includegraphics[width=\columnwidth]{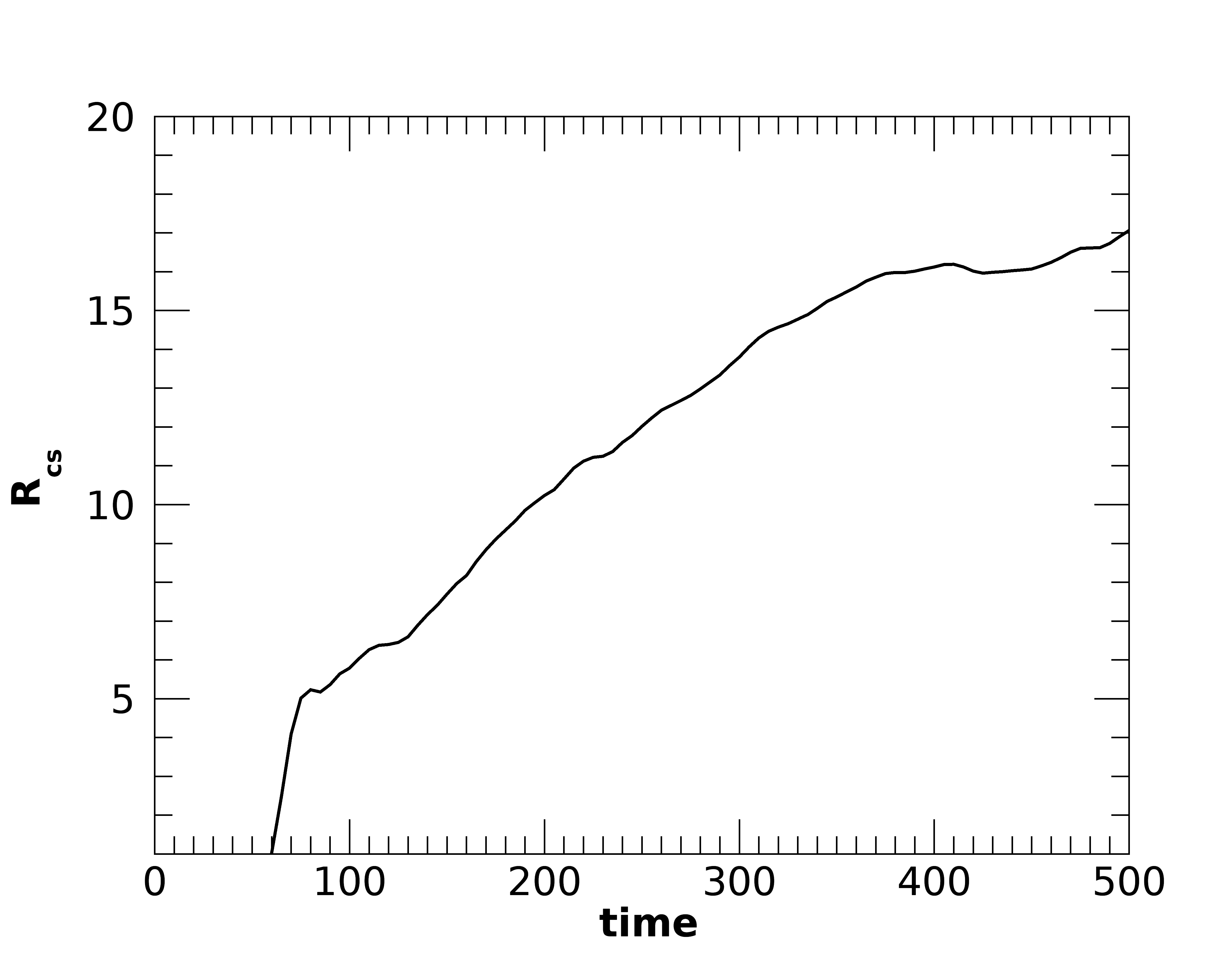}%
\caption{Plot of the radius $R_{cs}$, defined in equation (\ref{eq:rcs}), as a function of time computed using the value $s_{th}=0.2$.}
\label{fig:rwcs} 
\end{figure}

Finally we can see that the dissipation region, i.e., the region where we observe the formation of current sheets, tends to increase its size with time. In Fig. \ref{fig:rwcs} we show, as a function of time, the behavior of the quantity
\begin{equation}\label{eq:rcs}
R_{cs} =\frac{\int_{s > s_{th}} r j^2  dV}{\int_{s > s_{th}}  j^2  dV } \,,
\end{equation}
that represents an effective radius, weighted by $j^2$, of the dissipation region. We can observe a fast initial increase corresponding to the first development of the instability, during which $R_{cs}$ increases by a factor of 5, followed by a steady increase up to $R_{cs} \sim 15$, which then somewhat flattens at $t>400$ near the end of the simulation. Actually the bottom right panel of Fig. \ref{fig:current}, that depicts the current distribution at $t=500$, shows these current sheets, although much weaker, extending up to $r \sim 20$.

\begin{figure*}
\centering%
\includegraphics[width=\textwidth]{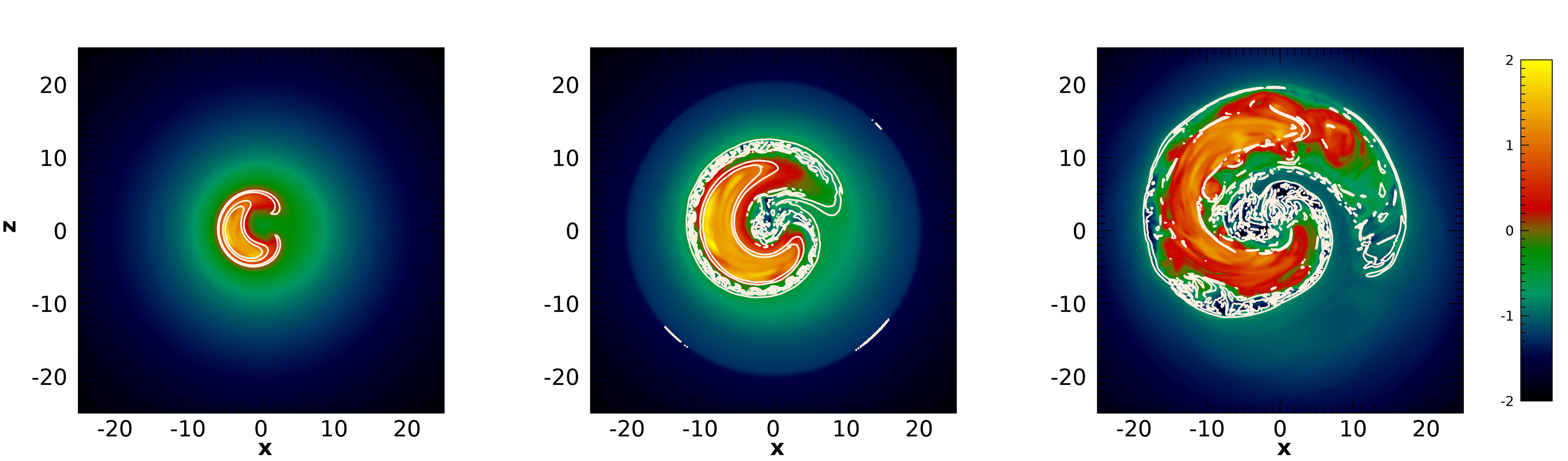}%
\caption{Two-dimensional cuts in the $x-y$ plane, at $z=8$, of the average magnetization with a superimposed contour plot of $s$ for $s=0.2$. The three panels refer to three different times: $t=66.6$ (left panel), $t=100$ (middle panel) and $t=166.6$ (right panel).}
\label{fig:sigma_cs} 
\end{figure*}

\begin{figure}
\centering%
\includegraphics[width=\columnwidth]{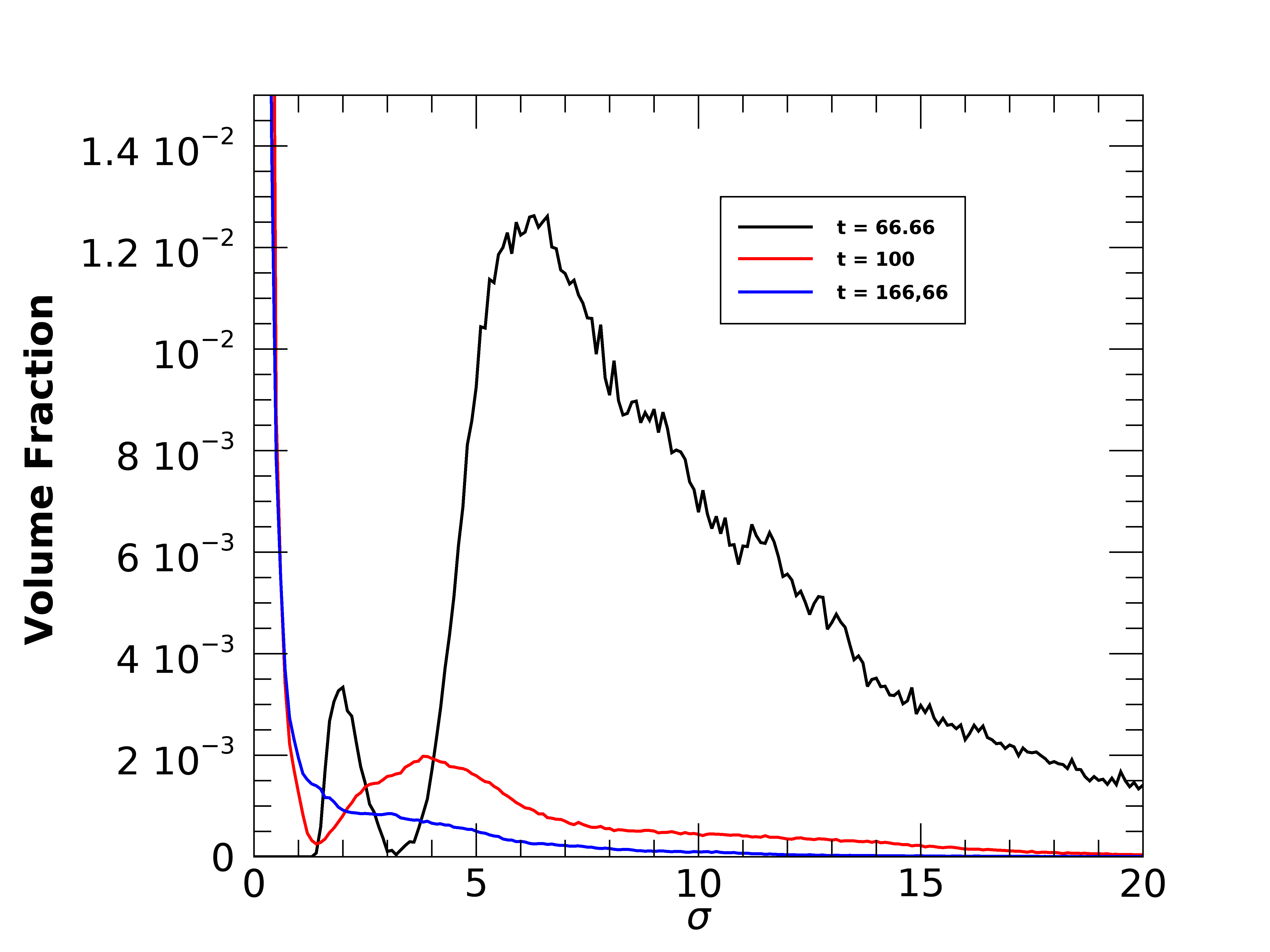}%
\caption{Plot of the volume fraction of current sheets with average magnetization $\sigma$ as a function of the latter. The three curves refer to three different times, as specified in the legend.}
\label{fig:sigma_histo} 
\end{figure}

An interesting quantity to analyse is the average magnetization at current sheets since PIC simulations  show that particle acceleration is more effective at higher magnetization. In Fig. \ref{fig:sigma_cs} we show the a two-dimensional cut in the $x-y$ plane (at $z=10$) of the average magnetization with superimposed white contours that show the location of current sheets. The average is computed over a cube with the side extending over 15 grid points. We consider the average rather than the punctual value in order to account for the magnetization values at the boundaries of the current sheets, which are those that are effectively considered in PIC simulations. The three panels refer to three different times: $t=66.6$ (right panel) just before the dissipation peak, $t=100$ (middle panel) just after the dissipation peak and $t=166.6$ at the beginning of the flat phase. The left panel shows current sheets that occupy the region of high magnetization, at later times however the current sheet position is found at the boundary of the high magnetization region. A more quantitative representation of the average magnetization of current sheets is given in Fig. \ref{fig:sigma_histo} in which we plot, for the same three times,  the volume fraction, with respect to the total current sheet volume, for a given value of the magnetization $\sigma$ as a function of $\sigma$. We can see that at $t=66.6$ we have a peak of the distribution at a value of $\sigma \sim 6$, a smaller peak at $\sigma \sim 1$ and a very extended tail reaching values $> 20$, At later times, instead, the distribution becomes concentrated at values $< 1$, for $t = 100$ we still observe a small peak at  $\sigma \sim 4$, that completely disappear at later times.

\subsubsection{Turbulence}
We have seen that, during the evolution, there are evidences of a transition  to a  turbulent state, as also pointed out in the recent simulations  by \citet{Bromberg19} and \citet{Davelaar2020}. In this section we will try to characterize in a more quantitative way when turbulence develops and its properties.

After the initial burst of the kink instability, at about $t=133.33$, the jet settles down into a turbulent state, characterized by nearly constant in time magnetic dissipation rate, as seen in Fig. \ref{fig:deth_j2}. In this state, the instability has already saturated. The structure of the turbulence in physical space is clearly seen in those panels of Figs. \ref{fig:comp} and \ref{fig:bfield} that correspond to the later times of the evolution. To better understand the saturation of the instability and characterize the properties of the ensuing turbulence, we analyse the spectral dynamics of perturbations. To this end, we perform Fourier transform of the variables in the $z$-direction along which the jet is periodic, 
\begin{equation}
\bar{g}(x,y,k_z,t)=\int_0^{L_z} g(x,y,z,t)\exp\left({-\rm i}k_z z \right)dz,
\end{equation}\label{eq:Fourier} 
where $g\equiv (A, {\bf B})$ with the first quantity in the brackets being the square root of the relativistic kinetic energy density, $A\equiv \sqrt{E_k}=\sqrt{\rho \gamma (\gamma-1)}$, and the second one the magnetic field. Due to the periodicity, as mentioned above, the longitudinal wavenumber $k_z$ is discrete, $k_z=2\pi n/L_z$, where the integer $n=0, \pm 1, \pm 2,...$. We define the kinetic, $\bar{E}_{kin}(k_z,t)=\int|\bar{A}|^2dxdy$, and magnetic $\bar{E}_{mag}(k_z,t)=(1/2)\int |\bar{\bf B}|^2dxdy$ spectra integrated in the entire $x-y$ plane of the domain. 

\begin{figure*}
\centering%
\includegraphics[width=\textwidth]{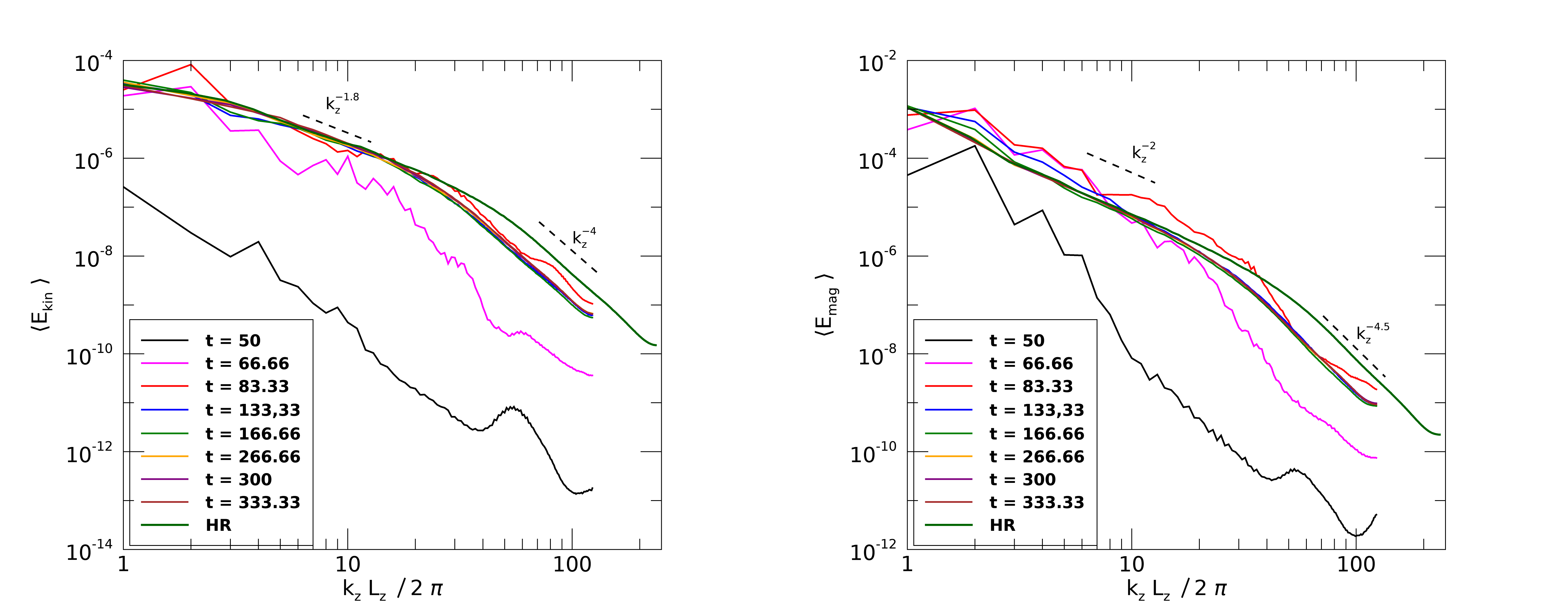}
\caption{Longitudinal spectra of the kinetic (left) and magnetic (right) energy densities at different times. Initially, the energies in the first few large scale $k_zL_z/2\pi=1,2$ kink modes grow faster with corresponding growth rates and consequently the spectra have a ``spiky'' irregular shape. Later, near the saturation time at about $t=83.33$, the nonlinearity comes into play, making the spectra smoother and smoother by transferring energy among modes. As a result, in the turbulent state, which sets in at about $t=133.33$, these spectra are essentially independent of time and at intermediate wavenumbers follow the power-law scaling $k_z^{-1.8}$ for the kinetic and $k_z^{-2}$ for the magnetic energies, whereas at large wavenumbers, where dissipation dominates, these spectra behave as $k_z^{-4}$ and $k_z^{-4.5}$, respectively. For comparison, in these panels we also show the kinetic and magnetic energy spectra for the high resolution run RefHR (thick green lines) in the turbulent state. These spectra well coincide with those of the lower resolution reference case at small and intermediate wavenumbers in the inertial range and further extend in this range to even higher wavenumbers.}
\label{fig:kin_mag_spectra} 
\end{figure*}

\begin{figure*}
\centering%
\includegraphics[width=\textwidth]{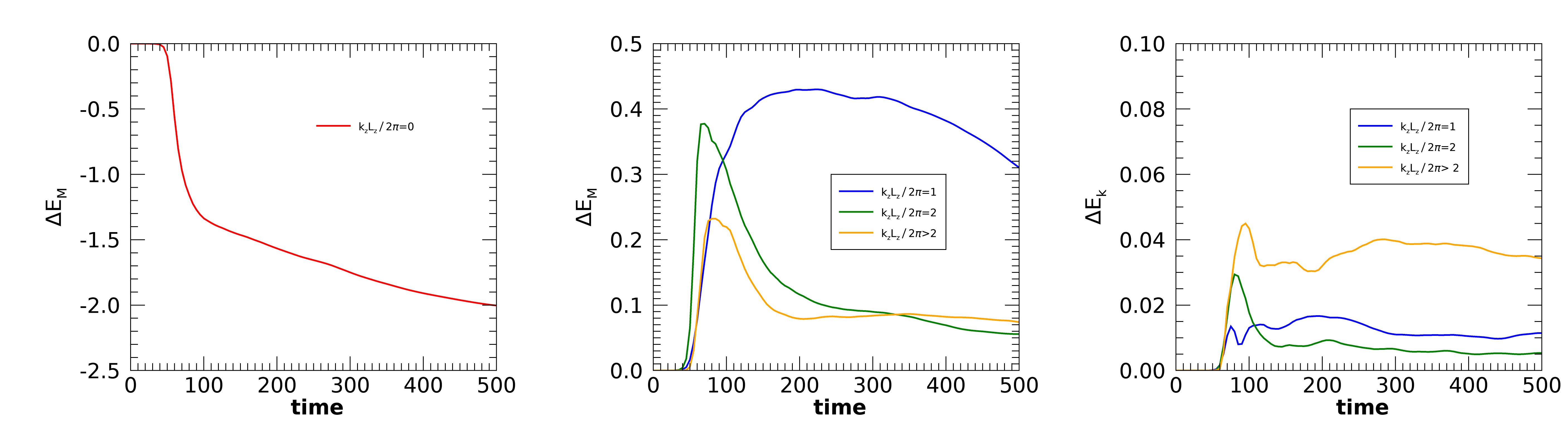}
\caption{Temporal evolution of the spectral energies integrated in the $x-y$ plane for different longitudinal wavenumbers $k_zL_z/2\pi=0~(red), 1~(blue), 2~(green)$ as well as the total values of these energies contained in larger wavenumbers $k_zL_z/2\pi > 2~(yellow)$. The left and middle panels show the magnetic energies, respectively, for the longitudinally uniform $k_z=0$ mode and higher $k_z$ modes, while the right panel shows the kinetic energies of $k_zL_z/2\pi \geq 1$ modes. The magnetic energy of the $k_z=0$ mode is shown with respect to its initial value. It is constant at the early linear stage, but decreases shortly afterwards during the whole evolution, supplying higher $k_zL_z/2\pi \geq 1$ modes. The kinetic and magnetic energies of the latter modes grow in the linear phase due to the kink instability and then remain nearly constant during the turbulent regime.} \label{fig:mode_energies_vs_t} 
\end{figure*}
 
\begin{figure*}
\centering%
\includegraphics[width=\textwidth]{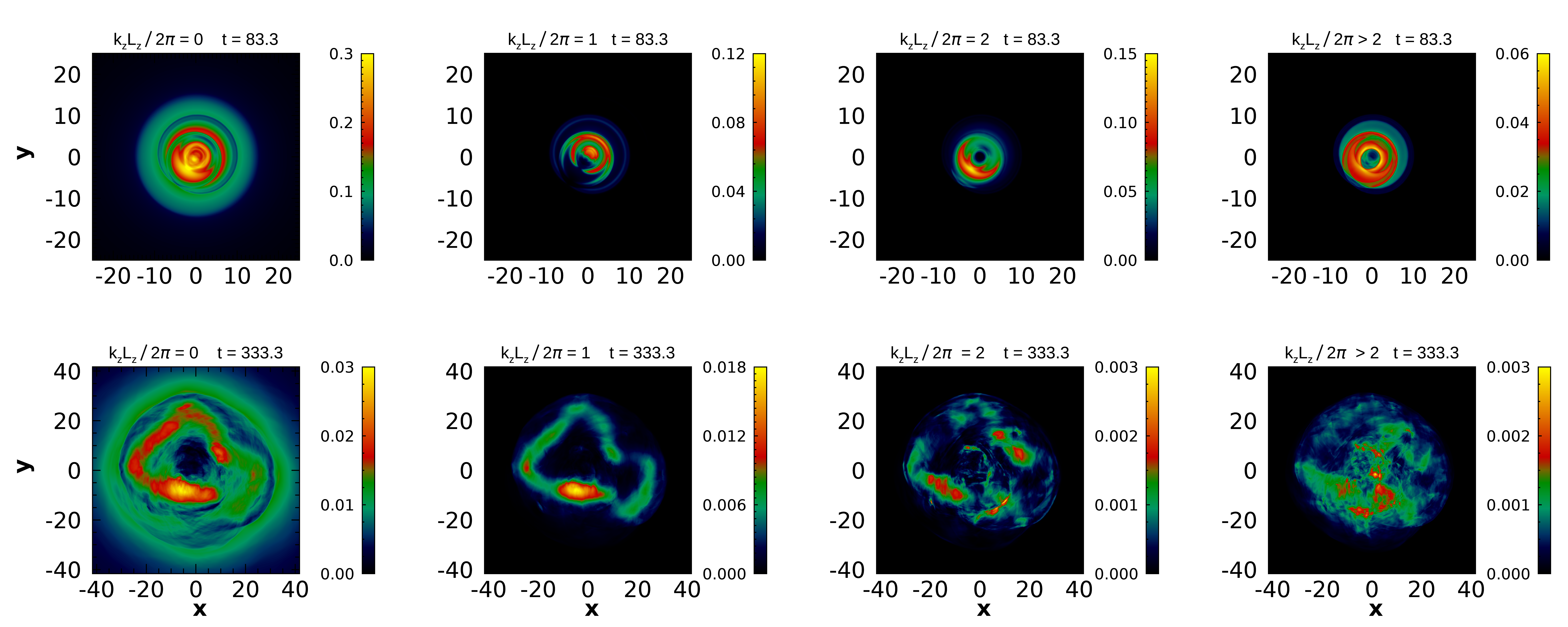}
\caption{Distribution of the spectral magnetic energy density, $\bar{E}_{mag}$ in the $x-y$ plane for $k_zL_z/2\pi=0, 1, 2$ modes as well as summed over all the modes with larger $k_zL_z/2\pi > 2$. The top row refers to the situation at $t=83.33$, when the growth of the kink instability has stopped. At this time, the magnetic field is still regular with dominant $m=0$ and $m=1$ azimuthal modes. The bottom row refers to the fully developed turbulence stage at $t=333.33$, when higher $k_zL_z/2\pi \geq 2$ modes have irregular turbulent structure, which is statistically nearly uniform across the jet, however, the first smallest $k_zL_z/2\pi=0, 1$ are still dominated by large-scale azimuthal modes.}
\label{fig:magenergy_slices} 
\end{figure*}

 Fig. \ref{fig:kin_mag_spectra} shows the temporal evolution of the kinetic and magnetic energy density spectra from $t=50$, corresponding to the growth of the kink instability at the initial linear stage (Fig. 4), up to the nonlinear stage at $t=333.33$, when the turbulence is already well developed. Fig. \ref{fig:mode_energies_vs_t} shows the time-development of the kinetic and magnetic energies for the individual large-scale $k_zL_z/2\pi=0, 1, 2$ modes as well as the total values of these energies contained in modes with higher wavenumbers $k_zL_z/2\pi > 2$. It is seen in these figures that at early, linear regime of the linear instability development, the kinetic and magnetic energies of these modes increase exponentially with different rates, with the mode $k_zL_z/2\pi=2$ being the most unstable (i.e., having the fastest growth rate) and hence dominant among other ones, as it is also apparent from the shape of the initial distortion of the jet in physical space (Fig. \ref{fig:comp}). The evolution of the longitudinally uniform $k_z=0$ mode is different from the other modes in that its magnetic energy is constant at the early stage and then monotonically decreases, first rapidly till about $t=100$, before the instability saturates, and then slower in the subsequent turbulent regime (Fig. \ref{fig:mode_energies_vs_t}). At the beginning of the evolution, the magnetic field of the $k_z=0$ mode corresponds to the initial field given by equations (6)-(8), which supports the kink instability and provides free energy necessary for its growth. As the kink-unstable, higher $k_z$ modes grow exponentially at the early stage, their feedback on the $k_z=0$ mode due to nonlinearity gradually increases as well, causing the large-scale magnetic field of the $k_z=0$ mode, which supports the instability, to decrease. Apart from the $k_z=0$ mode, the nonlinearity in the main equations redistributes the energy of the unstable modes among each other as well as to those higher wavenumbers, which are stable against the kink instability. Since the $k_z=0$ mode is the driver of the kink instability, weakening of this mode due to the nonlinear back-reaction of higher $k_z$ mode, in turn, leads to slowing down of the instability growth at later times compared to that in the linear stage. At the same time, the amplification of higher $k_z$ modes results in the increased intensity of the nonlinear transfers among these modes. A balance between these two processes -- the energy extraction by the unstable modes from the $k_z=0$ mode and its nonlinear redistribution -- is eventually reached at about $t=83.33$ implying the saturation of the instability, as it is seen in Fig. \ref{fig:mode_energies_vs_t}. After that the mode amplitudes no longer increase monotonically in time. In contrast to the ``spiky'' spectra at the initial linear growth stage, after the saturation, the kinetic and magnetic spectra have already become almost smooth and continuous, with all the $k_z$ modes being excited, and approaches a power-law form (Fig. \ref{fig:kin_mag_spectra}). However, at this time the jet is still not yet turbulent, being dominated by large-scale regular $m=0$ and $m=1$ azimuthal modes in the $x-y$ plane for all $k_z$, as seen in the upper row of Fig. \ref{fig:magenergy_slices}. 
 
 After a short transition phase, turbulence finally sets in at about $t=133.33$. From this moment, the amplitudes of all the nonzero $k_z$ modes are statistically quasi-steady in time except for that of the $k_z=0$ mode, whose magnetic field continues to decrease due to the nonlinear cascade towards higher $k_z$-modes (Fig. \ref{fig:mode_energies_vs_t}). Since this mode contributes most in the total magnetic energy, the latter similarly decreases with time (Fig. 4). It is evident in Fig. \ref{fig:kin_mag_spectra} that in the turbulent state, the kinetic and magnetic energy spectra are essentially independent of time and exhibit a typical feature of a turbulent spectrum -- power-law dependence on $k_z$, which at intermediate wavenumbers (inertial range) approximately follow $k_z^{-1.8}$ scaling for the kinetic and $k_z^{-2}$ for the magnetic spectra. At higher wavenumbers, where resistive dissipation occurs primarily in current sheets (Sec. 3.1.2), the kinetic and magnetic spectra follow steeper scalings $k_z^{-4}$ and $k_z^{-4.5}$, respectively. Thus, the large-scale magnetic field of the $k_z=0$ mode serves as a source, injecting energy in the turbulence by the kink instability. This is analogous to external forcing (often assumed to be distributed over a certain narrow  wavenumber range), which continuously feeds energy in the system, as used in forced turbulence cases, except that here this large-scale field slowly decays with time due to losing its energy to turbulence. This energy is transferred to higher $k_z$ modes due to nonlinearity and is ultimately dissipated at the smallest scales in the current sheets.  We confirmed these spectral properties of the turbulence in the high resolution case RefHR (thick green lines in Fig. \ref{fig:kin_mag_spectra}). Both the kinetic and magnetic energy spectra from the low and high resolution runs coincide at small and intermediate $k_z$ within the inertial range, which thus appears to be well represented already in the reference low resolution run. At high resolution, the same inertial range spectra further extend to higher $k_z$, shifting the dissipation range to even higher $k_z$ but not altering its shape.       
 
 The structure of the fully developed turbulent state in the $x-y$ plane for different $k_z$ modes at $t=333.33$ is shown in the bottom row of Fig. \ref{fig:magenergy_slices}. It is seen that higher $k_zL_z/2\pi \geq 2$ modes have irregular turbulent structure, which is nearly uniform across the jet, whereas the structure of smaller $k_zL_z/2\pi=0, 1$ modes is more regular, being dominated by large-scale azimuthal $m=0, 1$ modes. 

\subsection{Magnetization}
\begin{figure}
\centering%
\includegraphics[width=\columnwidth]{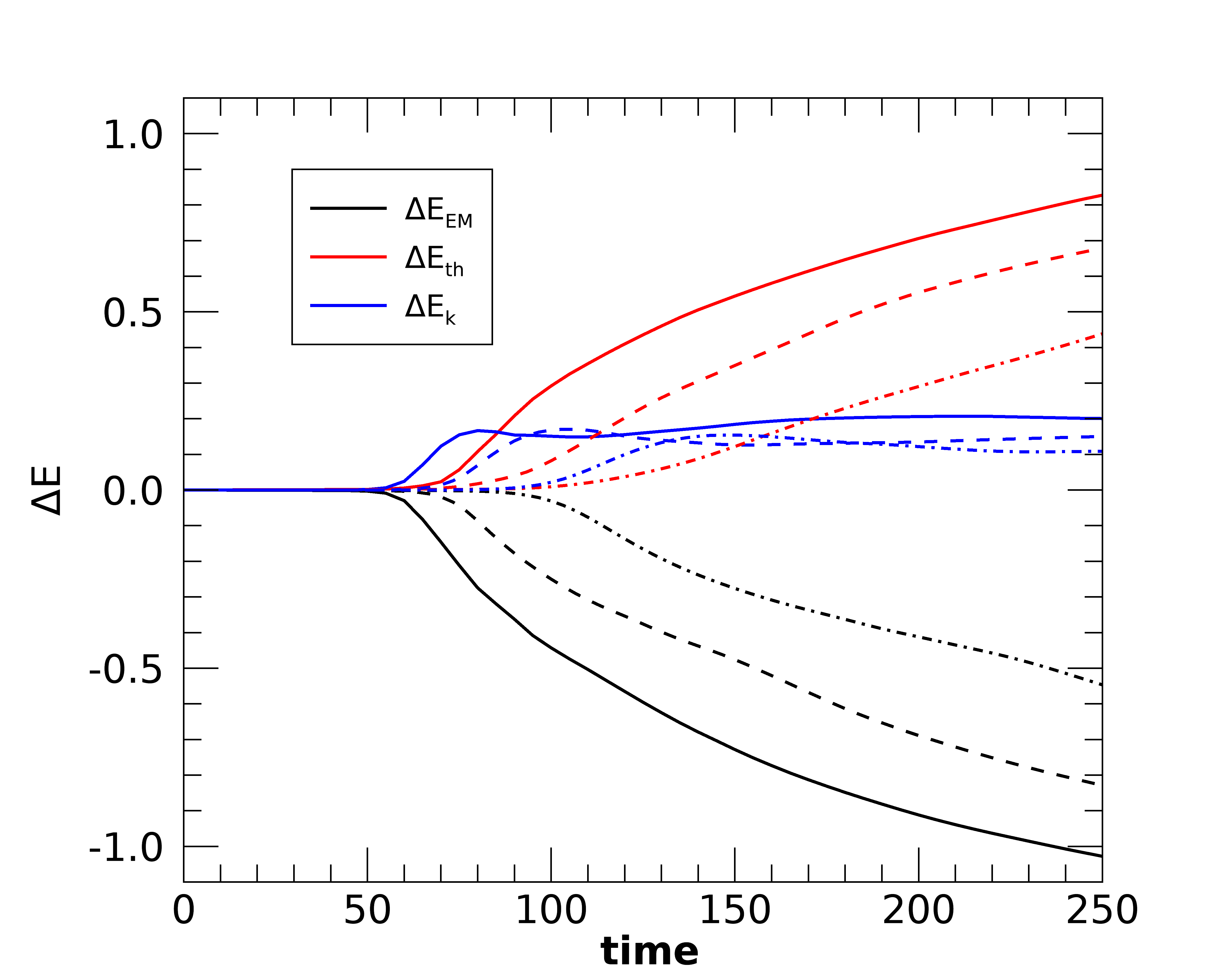}%
\caption{Plot of the variation of the three forms of energy (kinetic energy, thermal energy, electromagnetic energy), as a function of time. The energy variation are shown with respect to the value at $t=0$ and are normalized to the magnetic energy contained inside the magnetization radius $a$ at $t=0$. The solid curves refer to the case with $\sigma = 10$, the dashed curves to $\sigma = 2.5$ and the dashed-dotted curves to $\sigma = 1$. }
\label{fig:encomp_sigma} 
\end{figure}

\begin{figure*}
\centering%
\includegraphics[width=\textwidth]{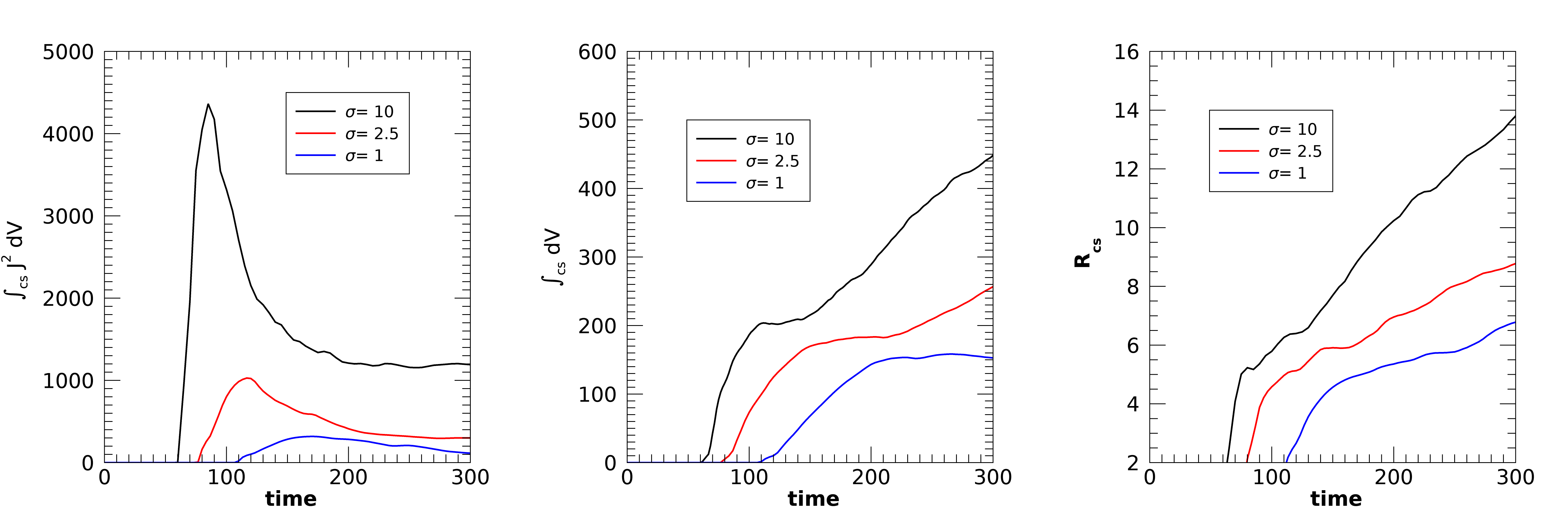}%
\caption{Left panel: plot of the integral of $j^2$ over the current sheets as a function of time. Middle panel: plot of the volume of current sheets as a function of time. Right panel: plot of the radius $R_{cs}$, defined in equation (\ref{eq:rcs}), as a function of time. The different curves in each panel refer to different values of the magnetization $\sigma$ as specified in the legends.}
\label{fig:cs_comp_sigma} 
\end{figure*}

Simulations Sigma2.5 and Sigma1 have the same configuration as the reference case, they differ only in the magnetization values which are, respectively, $\sigma=2.5$ and $\sigma = 1$. Consequently, we have values of the magnetic field strength that are lower by a factor of 2, in simulation Sigma2.5, and by a factor $\sim 3.1$, in simulation Sigma1. Fig. \ref{fig:encomp_sigma} shows the variation of the domain-averaged $E_M$ (black curves), $E_k$ (blue curves) and $E_{th}$ (red curves), with respect to their initial values, as functions of time, for the reference case (solid curves), the case Sigma2.5 (dashed curves) and Sigma1 (dash-dotted curves). The variations are normalized with respect to the initial value of the magnetic energy inside the magnetization radius. This normalization already accounts  for the differences in the available magnetic energy in these three cases, however, we observe an additional decrease of the normalized dissipated energy when we decrease the value of $\sigma$, implying a decrease in the efficiency of conversion of magnetic energy. The relative conversion into thermal or kinetic energies appears to be similar in the three cases. In Fig. \ref{fig:cs_comp_sigma} we plot, for the three cases, the integral of $j^2$ over all the current sheets (defined with $s_{th} = 0.2$) in the computational box as a function of time (left panel), the volume of the current sheets as a function of time (middle panel) and the radius $R_{cs}$ of the region occupied by current sheets defined in equation (\ref{eq:rcs}) (right panel). The decrease of the integral of the current with $\sigma$ appears to be somewhat higher than what could be expected from the simple scaling of the magnetic field strength. The middle panel shows that also the current sheet volume decreases with the magnetization. Both effects can explain the lowering of the magnetic energy conversion efficiency discussed above. In addition the region over which the current sheets extend shows a strong decrease of its radius with magnetization, as shown by the right panel of the figure. The ratio of the peak value to the value in the flat phase appears also to decrease with magnetization, implying a decreasing of the ratio of dissipation rates between the two phases. Additional differences between the three cases, that can be inferred from Figs. \ref{fig:encomp_sigma} and \ref{fig:cs_comp_sigma} are that the dissipation starts later, which is related to the decrease of the instability growth rate with magnetization \citep{Bodo13} and the widening of the peak in the integral of $j^2$, i.e. a longer duration of the peak phase.

\begin{figure}
\centering%
\includegraphics[width=\columnwidth]{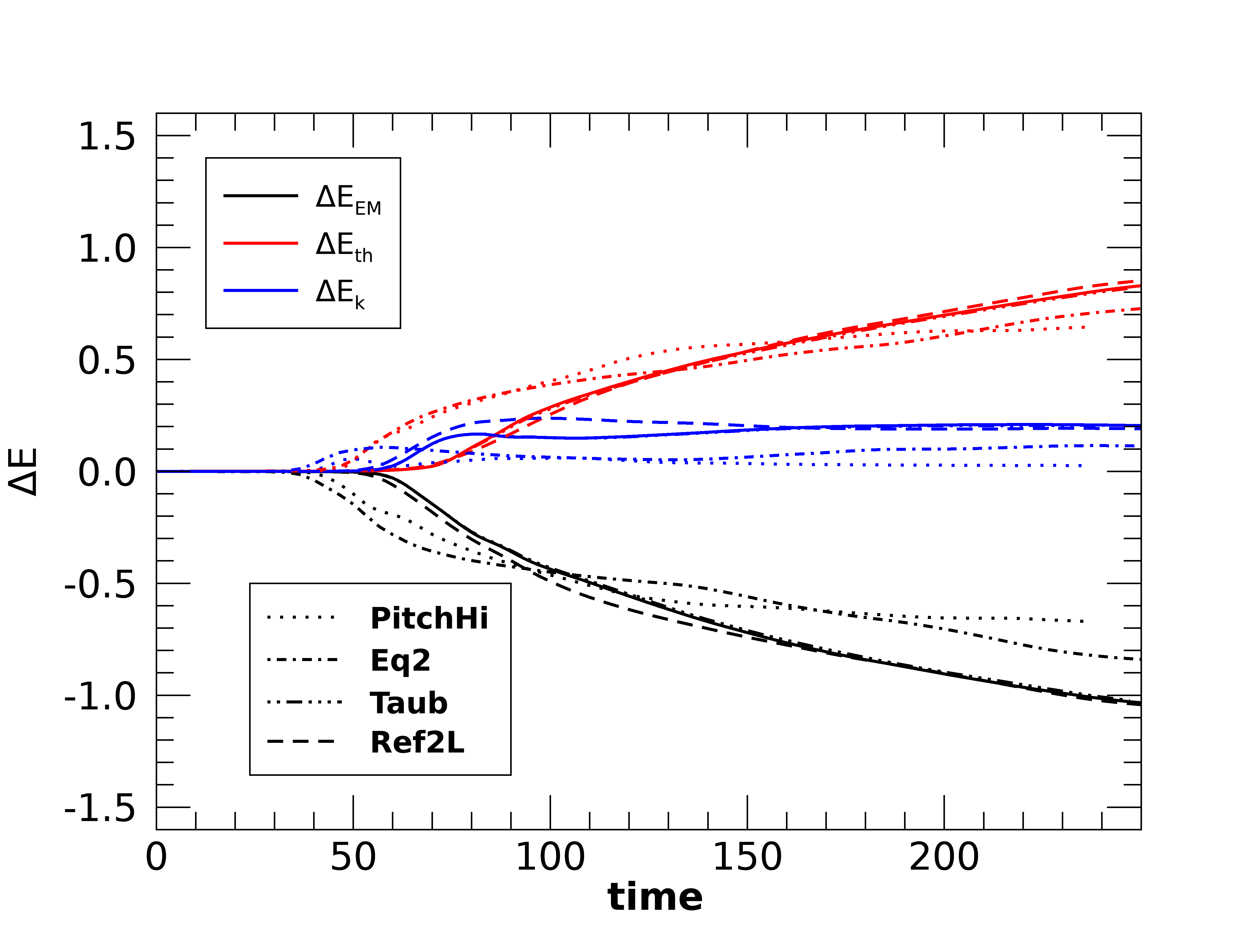}
\caption{Plot of the variation of the three forms of energy (kinetic energy, thermal energy, electromagnetic energy), as a function of time. The energy variations are shown with respect to the value at $t=0$ and are normalized to the magnetic energy contained inside the magnetization radius $a$ at $t=0$. The different curves refer to  different simulation setups as specified in the legend.}
\label{fig:setup_comp} 
\end{figure}

\begin{figure*}
\centering%
\includegraphics[width=\textwidth]{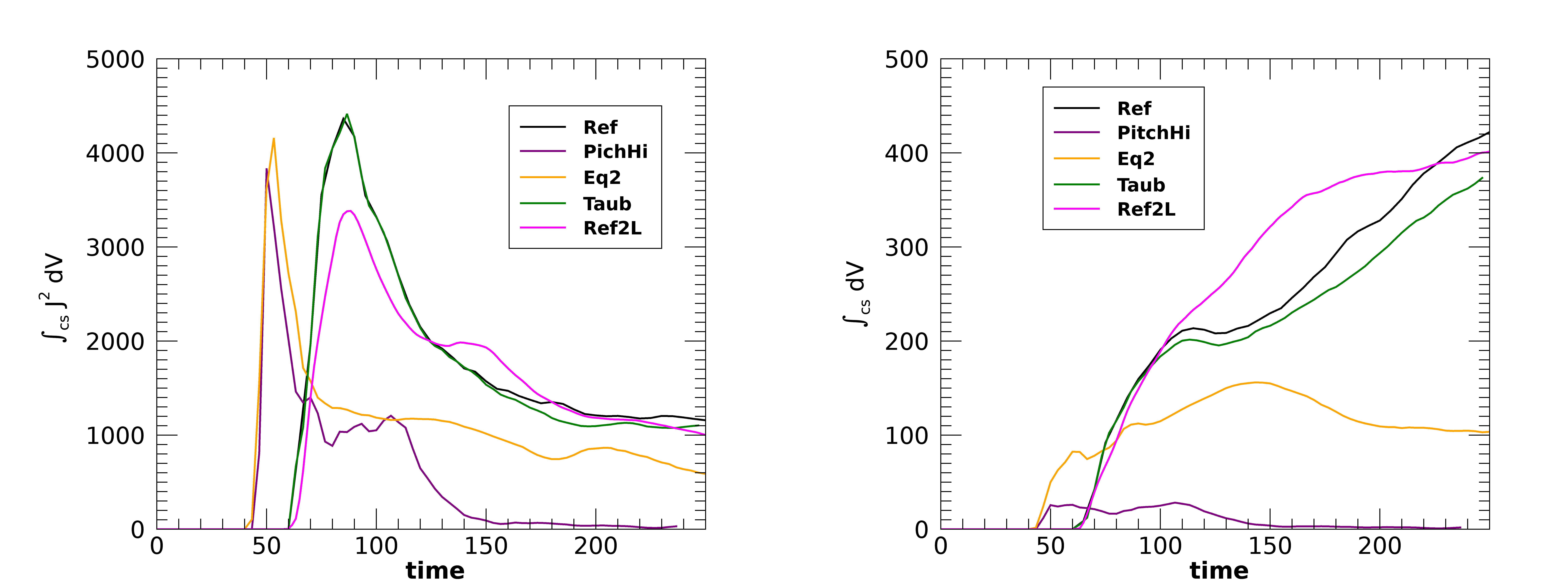}%
\caption{Left panel: plot of the integral of $j^2$ over the current sheets as a function of time. Middle panel: plot of the volume of current sheets as a function of time. Right panel: plot of the radius $R_{cs}$, defined in equation (\ref{eq:rcs}), as a function of time. The different curves in each panel refer to different simulation setups  as specified in the legends.}
\label{fig:setup_cs_comp} 
\end{figure*}

\subsection{Different  configurations}

In this Section we will discuss the differences induced by changing the configuration of the reference simulation in several different ways. We will analyse the consequences of changing the domain length by considering the results of simulation Ref2L in which we doubled the domain length. Different types of equilibria were used in simulations PitchHi and Eq2. More precisely, in simulation PitchHi the initial equilibrium has the same form of the Ref simulation, i.e. the form described in \citet{Bodo13}, but with a different value of the pitch parameter $P_c = -1.341$. As a result, there is a much larger axial field at large radial distances: in the case Ref, at large distances, $B_z = 1.7 \times 10^{-2}$, whereas in simulation PitchHi, the value of $B_z$ is $0.44$ (see also Fig. \ref{fig:equil}). The presence of an "external''  axial field, as we will see and as also shown by \citet{Bromberg19},  has  consequences for the instability evolution. In simulation Eq2, instead, we consider the equilibrium form introduced by \citet{Mizuno09} with an $\alpha$ parameter equal to $1.44$. Finally, in simulation Taub we used the Taub-Matthews equation of state \citep{Mignone05} instead of the constant $\Gamma$ equation of state used in the Ref simulation. During the course of the simulation we reach relativistic temperatures for which the use of constant $\Gamma = 5/3$ is not well suited, in simulation Taub we then consider the consequences of using a more appropriated equation of state.

For comparing the evolution of the different configurations, in Fig. \ref{fig:setup_comp} we plot the variations of the different form of energies as a function of time, while in Fig. \ref{fig:setup_cs_comp} we plot the integral of $j^2$ over the current sheets (left panel) and the current sheet volume (right panel) as functions of time. For reference and for making the comparison easier, in each plot we show  also the behavior of the Ref case. Before discussing in detail the changes in the evolution between the various cases, we note that, as it is evident from the figures, the instability growth rates may be different.

We start our discussion by examining the effects of the equation of state. In Fig. \ref{fig:setup_comp} the results for the Taub-Matthews equation of state are represented with the dashed-double dot curves that overlap almost exactly with those from the reference case. The same is true for the left panel in Fig. \ref{fig:setup_cs_comp}, while in the middle and right panels we observe somewhat larger differences at late times. From this we can then conclude that the effects of the equation of state are marginal. Also the case with a double length along the $z$ direction ($L_z = 33.33$, presented in Fig. \ref{fig:setup_comp} by the long dash curve) shows a behavior very similar to the Ref case. The main difference appears to be a somewhat lower value of the dissipation peak, which is likely related to the fact that the dominant mode, in the fist phase, is not the one with $k_z L_z / 2 \pi = 4$, as it can be expected from the doubling of the length, but the one with $k_z L_z / 2 \pi = 3$

More important is the role played by a different value of $P_c$, for which we have a non-negligible $B_z$ in the external medium. This case (PitchHi) is represented by the dashed curves in Fig. \ref{fig:setup_comp}, where it is seen that the efficiency of the conversion of magnetic energy is lower: at $t=250$ the magnetic energy converted in other forms is about $70\%$ of the reference case. In addition, the kinetic energy production is much lower than in the reference case (few percent compared to about $20\%$) and the energy variation curves tend to become flat for $t > 130$. This means that, at late times no energy conversion process is not taking place anymore. This is confirmed by  Fig. \ref{fig:setup_cs_comp}, which shows that the current sheets disappear, in fact the integral of $j^2$ over the current sheets almost vanishes for $t \geq 130$ and so does the volume. Notice however that \citet{Bromberg19} showed that in a similar configuration the amount of dissipation may also depend on the aspect ratio of the plasma column, since, for a longer column, more kink modes may grow and drive dissipation for a longer time.

In the case of the equilibrium Type 2, we also have a lower efficiency of the magnetic energy conversion and a lower kinetic energy production, although this effect is less pronounced with respect to the above PitchHi case. The dissipation rate at late times is lower than in the reference case, but, although showing a decrease with time, it does not drop to zero as in the previous case. The same is true for the current sheet volume. Besides, in both these last two cases, the extension of the dissipative region is reduced with respect to the Ref case. 

\begin{figure}
\centering%
\includegraphics[width=\columnwidth]{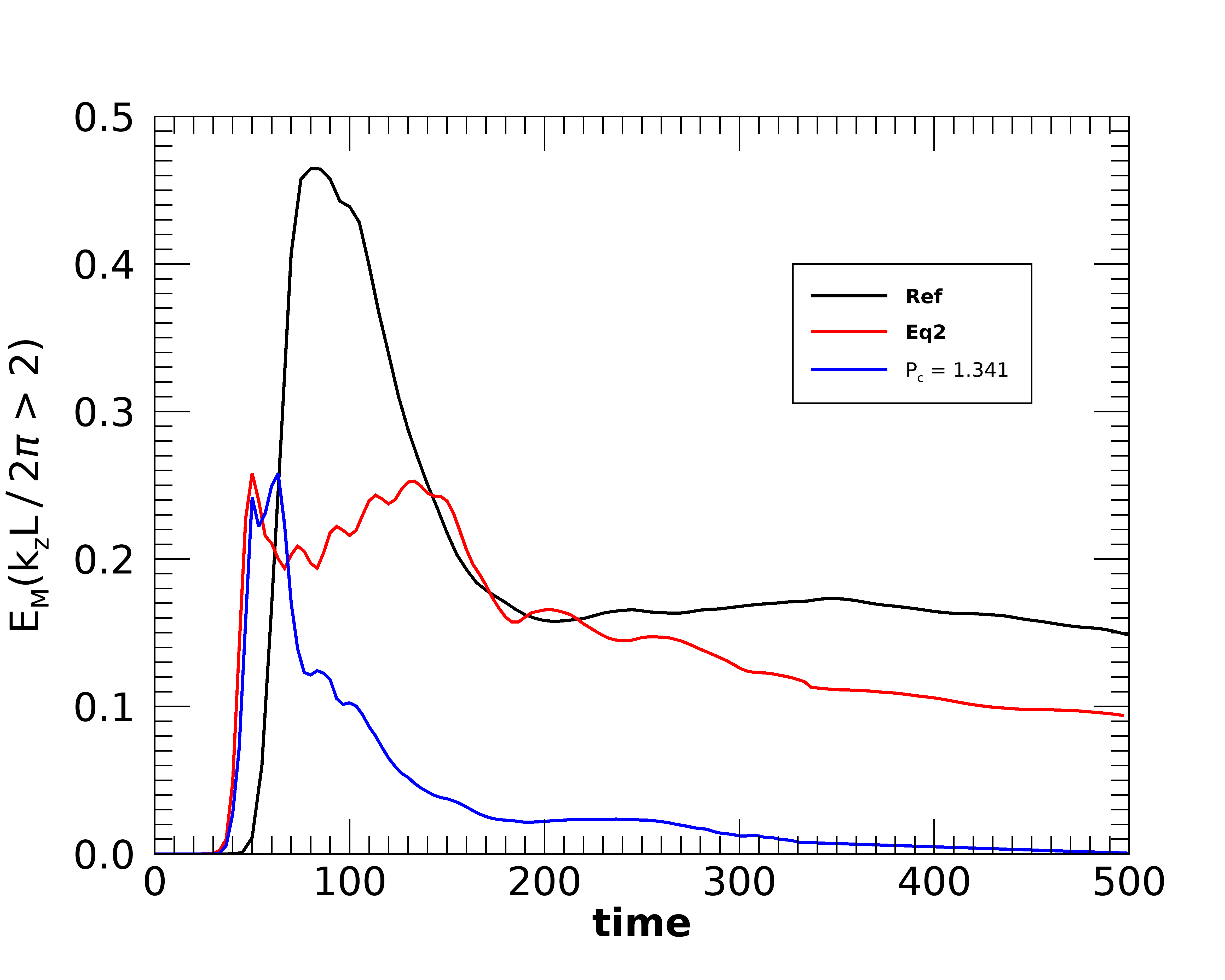}%
\caption{Plot of the magnetic energy integrated over all the modes with wavenumbers $k_z L_z / 2 \pi > 2$  as a function of time for the three cases shown in the legend. The energy is normalized with respect to the magnetic energy inside the magnetization radius at $t=0$.}
\label{fig:turb} 
\end{figure}

In both cases the lower magnetic energy dissipation is accompanied by a disappearance or a significant decrease of the energy in high wavenumber modes, as indicated by Fig. \ref{fig:turb}, which depicts the magnetic energy integrated over all the modes with wavenumbers $k_z L_z/ 2 \pi > 2$ as a function of time for the three cases Ref, Eq2 and PitchHi. We can see that, while in the Ref case the curve for $t > 200$ is almost flat with just a slight decrease towards the end, in the Eq2 case it shows a more pronounced decrease and in the PitchHi goes almost to zero. We can conclude that, in these two cases, turbulence is not any more sustained and either disappears or it is decaying. In both cases, the  structure of the magnetic field at late times is, on average, axisymmetric with typical pitch values for which the kink instability is absent and therefore there is no draining of energy towards higher wavenumbers.

\section{Summary and Discussion}
\label{sec:disc}
In this paper we analysed the evolution of current-driven kink instabilities in highly magnetized plasma columns, by means of numerical simulations carried out with the PLUTO code \citep{PLUTO}, considering different equilibrium configurations and focusing on the properties of the resulting dissipation of the magnetic energy. This analysis is relevant for high energy astrophysical sources, since the magnetic reconnection layers that form as a consequence of the instability evolution may provide a mechanism for the energization of non-thermal particles as shown by PIC simulations \citep[see e.g.][]{Sironi14, Guo14, Werner17, Petropoulou19}. The non-thermal relativistic particles may then originate  the observed  high energy emission. Understanding the dissipation characteristics is therefore important for the interpretation of these astrophysical objects \citep[see e.g.][]{Zhang16, Bodo21, Bromberg16}. 

Dissipation occurs through the formation of thin current sheets and its rate appear to depend only on the large scale characteristics of the flow and not on its dissipative properties. This  is demonstrated by a comparison of two cases of different resolution that, despite their different dissipative properties, show similar dissipation rates. In fact, in the higher resolution case the effect of a lower dissipation leads to the formation of thinner current sheets, while the total amount is essentially the same of the lower resolution case. This is a  point for which a more detailed analysis will be presented in a future paper.

We showed that dissipation proceeds in two stages, a peak, reached at the saturation of the instability, is followed by a longer phase of weaker strength. The duration and properties of these two phases depend on the initial equilibrium configuration and the physical parameters of the system. In the first phase, the instability evolution leads to the formation of an helicoidal current sheet  and the dissipation rate reaches a maximum. The helicoidal current sheet then starts to break in the central region and we observe the transition to a turbulent state that progressively extends outwards. In the latter case, dissipation has a lower rate and mainly proceeds instead in smaller-scale, more disordered current sheets. In this regard, we would like to mention that a similar nonlinear evolution of the kink instability in the main two -- first fast, via large-scale kink current sheet, and a later slower turbulence -- stages was observed and discussed by \cite{Bromberg19} and \cite{Davelaar2020} as well.

We also compared the behavior of different equilibrium configurations and found a dependence of the dissipative properties on the pitch profile and on the presence of an axial  magnetic field in the external regions of the plasma column. \citet{Bromberg19} already discussed the contrasting behaviours of different pitch profiles and showed that the case of decreasing pitch is more efficient in dissipating the magnetic energy than the case of increasing pitch. All the cases considered in this paper have decreasing profiles but different steepness. In fact, in the Ref case the pitch profile is steep and the magnetic field essentially vanishes at large radial distances from the column axis, in the Eq2 case the pitch has a shallower profile and, finally, the PitchHi case has a pitch profile very similar to the Ref case, but the axial magnetic field remains finite in the outside region. In addition, we also compared different magnetizations. The differences relate both to the total amounts of energy conversion and  to the duration of the turbulent phase. The Ref case is the one showing a larger fraction of magnetic energy converted in thermal and kinetic energies. Comparing different magnetizations, we showed that higher magnetization is more effective in dissipating energy not only because there is more energy available, but also due to the higher efficiency. In addition, we  analysed the evolution of the average magnetization in the regions around the current sheets since PIC simulations show that particle acceleration is more efficient for high magnetization \citep{Sironi14}. We found that, as the instability evolution proceeds, current sheets are found in regions with progressively lower magnetization and we therefore expect a decrease in the particle energization process, which could translates in an effective shortening of the radiative peak associated to the dissipation peak. 

The behaviour in the second turbulent phase is also different in the different cases. In the Ref case, the turbulence persists up to the final time of the simulation ($t_f = 500$), showing only a slight decay towards the end of the simulation. In the Eq2 case the turbulence shows a stronger decay, while in the PitchHi the turbulence dies very soon. These behaviours may be related to the attainment of a relaxed state discussed by \citet{Bromberg19} and \citet{Browning08}. In the Ref case, the configuration reached at the end of the simulation is still far from a relaxed state and may still feed energy into the turbulence, while in the two other cases the final configurations are closer to relaxation when the energy transfer to turbulence has stopped. This may be important in connection with the possible energization of non-thermal particles, since following the first phase, in which the energization can occur essentially through the reconnection layers, in the second phase, turbulence may also provide an alternative path for particle acceleration. This can also have an impact on the polarization properties of the emitted radiation: in the first phase we expect a high polarization fraction related to the well ordered structure of the magnetic field \citep{Bodo21}, in the second turbulent phase, instead, we could expect a lower polarization fraction, as a result of the more disordered magnetic field structure. 

These results refer to a static plasma column, clearly it would be important to investigate how they are possibly modified by the presence of a velocity shear between the jet and the ambient medium. This will be investigated in a follow-up paper.

\section*{Acknowledgments}
This project has received funding from the European Union's Horizon
2020 research and innovation programme under the ERC Advanced Grant
Agreement No. 787544 and from the Shota Rustaveli National
Science Foundation of Georgia (SRNSFG, grant No. FR17-107). We acknowledge the computing centre  of Cineca and INAF, under the coordination of the "Accordo Quadro MoU INAF-CINECA 2017-2021", for the availability of computing resources and support.
GB and PR acknowledge contribution from the grant INAF CTA–
SKA “Probing particle acceleration and $\gamma$-ray propagation with
CTA and its precursors” and the INAF Main Stream project “High-
energy extragalactic astrophysics: toward the Cherenkov Telescope
Array”.

\section{Data availability}
Data available on request

\bibliographystyle{mn2e}
\bibliography{main.bib}

\end{document}